\documentclass{aa}

\usepackage[varg]{txfonts}
\usepackage{natbib}
\usepackage{color}
\usepackage{subcaption}
\usepackage{stackengine}

\usepackage{xspace}
\usepackage{ulem}

\bibpunct{(}{)}{;}{a}{}{,}

\newcommand{\subfigcolorlabel}[3]{
    \begin{subfigure}[t]{0.5\textwidth}
        \topinset{\textcolor{#3}{#2}}{
            \includegraphics[height=0.37\textheight,keepaspectratio]{#1}
        }{0.12\textwidth}{-0.3\textwidth}
    \end{subfigure}}

\newcommand{\subfigfivecolorlabel}[3]{
    \begin{subfigure}[t]{0.47\textwidth}
        \topinset{\textcolor{#3}{#2}}{
            \includegraphics[height=0.39\textheight,keepaspectratio]{#1}
        }{0.15\textwidth}{-0.3\textwidth}
    \end{subfigure}}

\newcommand{\subfigcolorrightlabel}[3]{
    \begin{subfigure}[t]{0.5\textwidth}
        \topinset{\textcolor{#3}{#2}}{
            \includegraphics[width=1.0\textwidth]{#1}
        }{0.1\textwidth}{0.287\textwidth}
    \end{subfigure}}

\newcommand{\subfigcolorrightlabelnew}[3]{
    \begin{subfigure}[t]{0.5\textwidth}
        \topinset{\textcolor{#3}{#2}}{
            \includegraphics[width=1.0\textwidth]{#1}
        }{0.15\textwidth}{0.287\textwidth}
    \end{subfigure}}

\newcommand{\subfigcolorsizelabel}[4]{
    \begin{subfigure}[t]{#4\textwidth}
        \topinset{\textcolor{#3}{#2}}{
            \includegraphics[height=1.7\textwidth]{#1}
        }{0.08\textwidth}{-0.15\textwidth}
    \end{subfigure}}

\newcommand{\nnhp}{\mbox{N$_2$H$^+$}\xspace}

\newcommand{\nhhh}{\mbox{NH$_3$}\xspace}
\newcommand{\sio}{\mbox{SiO}\xspace}
\newcommand{\HII}{\mbox{H\thinspace{\sc ii}}\xspace}

\newcommand\micron{\mbox{$\mu$m}\xspace}
\newcommand\um{\micron}
\newcommand\arcdeg{\mbox{$^\circ$}\xspace}

\newcommand{\herschel}{\textit{Herschel}\xspace}

\newcommand{\Tbg}{\mbox{$T_{\mathrm{bg}}$}\xspace}

\newcommand{\Trot}{\mbox{$T_{\mathrm{rot}}$}\xspace}
\newcommand{\Tkin}{\mbox{$T_{\mathrm{kin}}$}\xspace}

\newcommand{\Tex}{\mbox{$T_{\mathrm{ex}}$}\xspace}
\newcommand{\Ntot}{\mbox{$N_{\mathrm{tot}}$}\xspace}
\newcommand{\voff}{\mbox{$v_{\mathrm{off}}$}\xspace}
\newcommand{\sig}{\mbox{$\sigma$}\xspace}
\newcommand{\kmspc}{\mbox{$\mathrm{km~s^{-1}~pc^{-1}}$}\xspace}
\newcommand{\kms}{\mbox{$\mathrm{km~s^{-1}}$}\xspace}
\newcommand{\ms}{\mbox{$\mathrm{m~s^{-1}}$}\xspace}
\newcommand{\vlsr}{\mbox{$v_{\mathrm{lsr}}$}\xspace}
\newcommand{\Ms}{\mbox{$M_{\sun}$}\xspace}
\newcommand{\Mspc}{\mbox{$M_{\sun}~\mathrm{pc}^{-1}$}\xspace}
\newcommand{\asec}{\mbox{\arcsec}\xspace}

\begin{document}
\title{Temperature structure and kinematics of the IRDC G035.39--00.33}

\author{Vlas Sokolov\inst{1}
  \and Ke Wang\inst{2}
  \and Jaime E. Pineda\inst{1}
  \and Paola Caselli\inst{1}
  \and Jonathan D. Henshaw\inst{3}
  \and Jonathan C. Tan\inst{4,5}
  \and Francesco Fontani\inst{6}
  \and Izaskun Jim\'enez-Serra\inst{7}
  \and Wanggi Lim\inst{4}
  }

\institute{Max Planck Institute for Extraterrestrial Physics, Gie{\ss}enbachstra{\ss}se 1, 85748 Garching bei M{\"u}nchen, Germany
  \and European Southern Observatory, Karl-Schwarzschild-Str. 2, D-85748, Garching bei M{\"u}nchen, Germany
  \and Astrophysics Research Institute, Liverpool John Moores University, Liverpool, L3 5RF, UK
  \and Department of Astronomy, University of Florida, Gainesville, FL, 32611, USA
  \and Department of Physics, University of Florida, Gainesville, FL, 32611, USA
  \and INAF-Osservatorio Astrofisico di Arcetri, Largo E. Fermi 5, I-50125 Firenze, Italy
  \and School of Physics and Astronomy, Queen Mary University of London, Mile End Road, London E1 4NS, UK
  }

\abstract
{}
{Infrared dark clouds represent the earliest stages of high-mass star formation. Detailed observations of their physical conditions on all physical scales are required to improve our understanding of their role in fueling star formation.}
{We investigate the large-scale structure of the IRDC G035.39--00.33, probing the dense gas with the classical ammonia thermometer. This allows us to put reliable constraints on the temperature of the extended, pc-scale dense gas reservoir and to probe the magnitude of its non-thermal motions. Available far-infrared observations can be used in tandem with the observed ammonia emission to estimate the total gas mass contained in G035.39--00.33.}
{
We identify a main velocity component as a prominent filament, manifested as an ammonia emission intensity ridge spanning more than 6 pc, consistent with the previous studies on the Northern part of the cloud.
A number of additional line-of-sight components are found, and
a large scale, linear velocity gradient of ${\sim}0.2$ \kmspc is found along the ridge of the IRDC.
In contrast to the dust temperature map, an ammonia-derived kinetic temperature map, presented for the entirety of the cloud, reveals local temperature enhancements towards the massive protostellar cores.
We show that without properly accounting for the line of sight contamination, the dust temperature is 2-3 K larger than the gas temperature measured with \nhhh.
}
{While both the large scale kinematics and temperature structure are consistent with that of starless dark filaments, the kinetic gas temperature profile on smaller scales is suggestive of tracing the heating mechanism coincident with the locations of massive protostellar cores.}

\keywords{ISM: kinematics and dynamics -- ISM: clouds -- stars: formation -- ISM: individual objects: G035.39-00.33}
\maketitle

\section{Introduction}
\label{sec:intro}

Massive ($M_{\star}> 8~M_{\odot}$) stars dominate their environments through powerful stellar winds, ionizing radiation, and their decisive role in driving the turbulence and enriching the chemical complexity of the interstellar medium.
Despite the importance massive stars play in their host galaxies,
understanding the earliest phases of their formation is still an ongoing effort \citep[e.g.][for a review of the subject]{tan+2014}.

The majority
of massive stars are not formed in isolation. A large fraction of all stars are born within Giant Molecular Clouds (GMCs) \citep{mckee+ostriker2007}, massive ($> 10^4~M_{\odot}$), often filamentary, molecular structures that span across dozens of parsecs and are thought to be responsible for the bulk of Galactic star formation.
GMCs have been found to be highly sub-structured, and with the advance of mid- and far-infrared imaging instruments, the ubiquity of filamentary structure in star-forming molecular clouds became apparent \citep{molinari+2010, andre+2010}. As the filaments assemble their mass, the densest filaments in star-forming clouds are thought to become gravitationally unstable, fragmenting further into protostellar cores.

Often tracing the highest density regions of GMCs \citep[e.g.][]{schneider+2015}, Infrared Dark Clouds (IRDCs) have been used as testing grounds for the earliest stages of massive star and cluster formation theories since the late 1990's. First identified as dark features obscuring the bright Galactic background \citep{perault+1996, egan+1998}, IRDCs soon became recognized as the most promising candidates for harboring the long-sought initial conditions of high-mass star forming regions \citep[e.g.][]{rathborne+2006}.
Subsequent far-infrared, submillimetre, and radio band observations revealed a high degree of fragmentation in these clouds, with dense and massive cores exhibiting a variety of star formation stages: from prestellar, dark, cold, and quiescent cores to active, infrared-bright and chemically rich substructures with embedded sources driving outflows and \HII regions \citep[e.g.,][]{beuther+2005, pillai+2006-nh3, chambers+2009, sanhueza+2012, wang+2011, wang+2014}.

G035.39--00.33 (hereafter G035.39)
is a cold \citep[15 -- 17 K,][]{nguyen_luong+2011},
massive \citep[$\sim$16700 M$_{\sun}$,][]{kainulainen+tan2013} IRDC located 2.9 kpc away in the W48 molecular complex \citep{simon+2006}. Its highly filamentary structure appears as an extinction feature up to 70\micron, and the cloud harbors a number of dense cores \citep{butler+tan2012}.

Previous single-dish radio and far-infrared studies of G035.39 describing the large, pc-scale, gas reservoir suggest that the bulk of the cloud material in the IRDC
represents the typical chemical properties of cold and dense gas, namely high CO depletion \citep{hernandez+2011, hernandez+2012, jimenez-serra+2014} and high values of deuteration \citep{barnes+2016}.
Furthermore, the dust temperature maps in \cite{nguyen_luong+2011}, derived from \herschel photometric maps, show a monotonic decrease in dust temperatures from the edges of the IRDC to its innermost regions, where most of the massive protostellar cores are, with no apparent heating signatures of the embedded protostars in G035.39.

\begin{figure}
    \centering
    \makebox[\textwidth][l]{\includegraphics[width=0.5\textwidth]{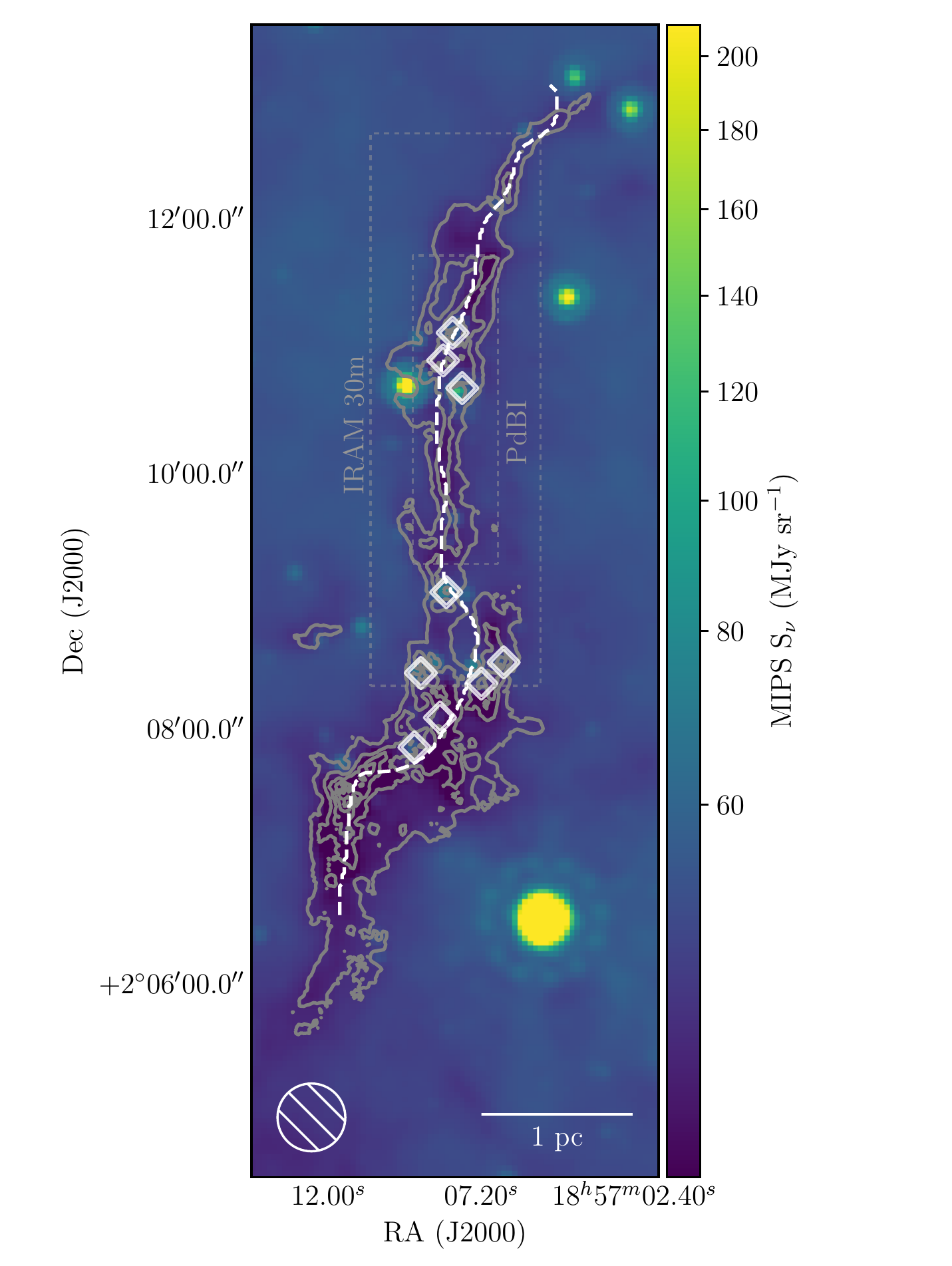}}
    \caption{MIPS 24\micron emission overlaid with infrared extinction contours \citep{kainulainen+tan2013} starting from $\mathrm{A_V}=25$ mag and progressing inwards in steps of 20 mag. Massive dense \herschel cores from \cite{nguyen_luong+2011} are marked with white diamonds. The white dashed line indicates the peak intensity ridge of \nhhh (1, 1) emission (see \S\ref{sec:gbtobs}). The dashed grey rectangles illustrate the extent of IRAM 30m (outer box) and PdBI (inner box) \nnhp maps \citep{henshaw+2013, henshaw+2014}, while the extent of the whole figure corresponds to the area mapped with the GBT.
    }
    \label{fig:overview}
\end{figure}

\cite{jimenez-serra+2010} observed a widespread, pc-scale \sio emission as a mixture of broad and narrow components, finding it consistent with being a remnant of a large-scale shock, possibly associated with the IRDC formation process. Follow up observations
revealed the northern part of the cloud to possess complex, multicomponent kinematics, with the velocity components interacting dynamically with the massive cores \citep{henshaw+2013, henshaw+2014, jimenez-serra+2014}.
Alternatively, the origin of the \sio emission across the cloud could be attributed to outflow activities of undetected embedded protostars \citep{jimenez-serra+2010}.
Indeed, \cite{nguyen_luong+2011} find 70\micron \herschel sources in 13 massive dense cores (20-50 M$_{\sun}$, MDCs) in G035.39, which indicates that these cores are potentially forming high-mass stars (white diamonds on Fig. \ref{fig:overview}). While the \herschel sources are likely to be responsible for the observed broad component \sio emission in G035.39, attributing its narrow component to embedded protostars would require existence of an undetected population of low-mass protostars across the cloud.

By comparing the C$^{18}$O line emission with the mass surface density map obtained from extinction mapping,
\cite{hernandez+2012} concluded that the denser part of the cloud is consistent with being in virial equilibrium. 
Follow-up observations of G035.39 with the Plateau de Bure interferometer (PdBI) have resolved the kinematics of the cloud into distinct sub-virial, velocity coherent structures that hint at their dynamical interaction with an embedded protocluster \citep{henshaw+2014, henshaw+2016_pdbi}.
The cores in the continuum substructure appear to be intertwined in a network of independent filamentary structures \citep{henshaw+2017_alma} and are likely to collapse
without additional support from magnetic fields (\citeauthor{henshaw+2016_pdbi} \citeyear{henshaw+2016_pdbi}, see also \citeauthor{tan+2013} \citeyear{tan+2013}).

Despite the wealth of observations collected for this IRDC, its gas temperature structure has never been mapped before.
This paper intends to
establish a coherent picture of the physical conditions of dense gas across the whole extent of the IRDC using observations of two ammonia inversion lines obtained with a high spectral resolution.
Figure \ref{fig:overview} shows the infrared extinction morphology of G035.39 and marks the portion of the cloud studied by \cite{henshaw+2013, henshaw+2014}. The overall field of view of the figure shows the extent of observations that will be presented in this study. For comparison to the previous body of work, we will refer to the region north of $\mathrm{\delta(J2000) = +2\arcdeg08\arcmin45\arcsec}$, approximately corresponding to the extent of previous IRAM 30m and PdBI studies, as G035.39-N.

We discuss observations conducted and available data used in \S\ref{sec:obs}.
Dust temperature, gas column density, cloud kinematics, and ammonia abundance are presented in \S\ref{sec:results}.
We compare the gas and dust temperatures, and discuss the stability of the cloud, in \S\ref{sec:discussion}.
We summarize our findings in \S\ref{sec:conclusions}.

\section{Observations}
\label{sec:obs}
\subsection{GBT observations}
\label{sec:gbtobs}
Robert C. Byrd Green Bank Telescope (GBT) observations of G035.39 were carried out on 2010 Feb 27th and 28th (project GBT/10A-067, PI: Ke Wang). G035.39 was covered in Nyquist sampled on-the-fly (OTF) maps scanning along RA and Dec. The K-band receiver was equipped with two beams (B3 and B4) separated by 178.8 arcsec in azimuth. We used the frequency switching mode, with $\pm$2.5 MHz switch for the two signal states, in a 12.5 MHz band with a channel width of 3.05 kHz in dual polarization mode. The band covered \nhhh (1,1) and (2,2) inversion lines, CCS (2-1) rotational transition, and NH$_2$D ($4_{14}$-$4_{04}$), the latter being undetected. The weather conditions at the GBT site were stable during the observing run, with pointing accuracy resulting from winds alone estimated to be below 3\asec. The system temperatures were at 38-45 K during the first day and between 58-78 K during the second day.

To estimate the absolute flux of the observed emission, we have performed nodding observations of a quasar 3C48. For each beam, polarization, and spectral window states, the data were then reduced in GBTIDL v2.10.1 to yield main beam temperatures for each state. Using a spectral flux density model from \cite{ott+1994}, we derived the predicted flux for 3C48. When compared to reduced nodding observations, the model is consistent within 10\% for ammonia spectral windows. The off-source beam B4, however, was found to be mismatched by 50-90\% in integrated intensity when compared to the on-source beam, and was consequently removed from the following discussion. As the area mapped by B4 was mainly off the bulk of G035.39, the decision to drop the beam from a subsequent discussion does not affect the results of this study.

The OTF observations of G035.39 were calibrated via the GBT pipeline\footnote{\url{http://gbt-pipeline.readthedocs.io/en/latest/}} \citep{masters+2011} into main beam temperature units and imaged with the GBT Ammonia Survey
(GAS, Friesen, Pineda et al. \citeyear{friesen+pineda+2017})
pipeline\footnote{\url{http://gas.readthedocs.io/en/latest/}},
modified to address spectral setup differences. The full width at half maximum of the GBT beam at \nhhh (1,1) line frequency (32\asec) was used to set the pixel size for all spectral line cubes, at three pixels per beam. The final spectral resolution of the data is 38.6 \ms.

\subsection{Herschel public data}
\label{sec:herschelobs}

As mentioned in \S\ref{sec:intro}, the entire extent of G035.39 has been studied before with the \herschel Space Observatory \citep{pilbratt+2010} by \cite{nguyen_luong+2011}. To quantify the difference in gas- and dust-derived properties, we use the available G035.39 data from the \herschel infrared Galactic Plane Survey \citep[Hi-GAL,][]{molinari+2010}.

The data products used in this study, photometric maps from PACS \citep{poglitsch+2010} and SPIRE \citep{griffin+2010} cameras, were downloaded from the image server of the first public Hi-GAL data release \citep[DR1,][]{molinari+2016}, observation id's 1342219631 and 1342219630. The DR1 data underwent processing by the ROMAGAL pipeline \citep{traficante+2011}, and in particular, had its absolute levels of emission calibrated. This allows us to directly put constraints on the properties of dust emission.

\section{Results}
\label{sec:results}

\subsection{Overview of the data}
\label{sec:dataoverview}

\begin{figure*}[h!]
    \subfigcolorsizelabel{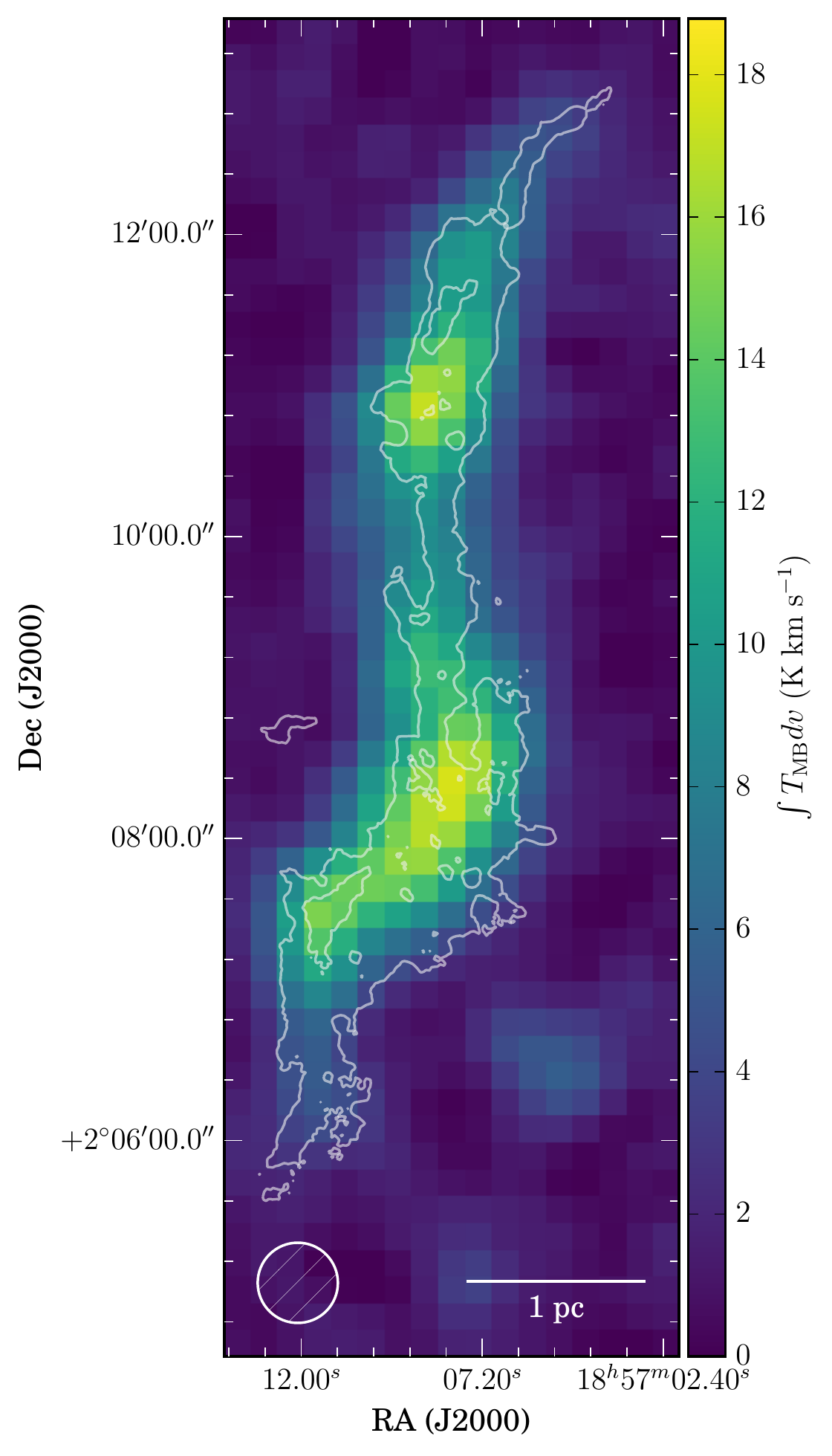}{(a)}{white}{0.33}
    \subfigcolorsizelabel{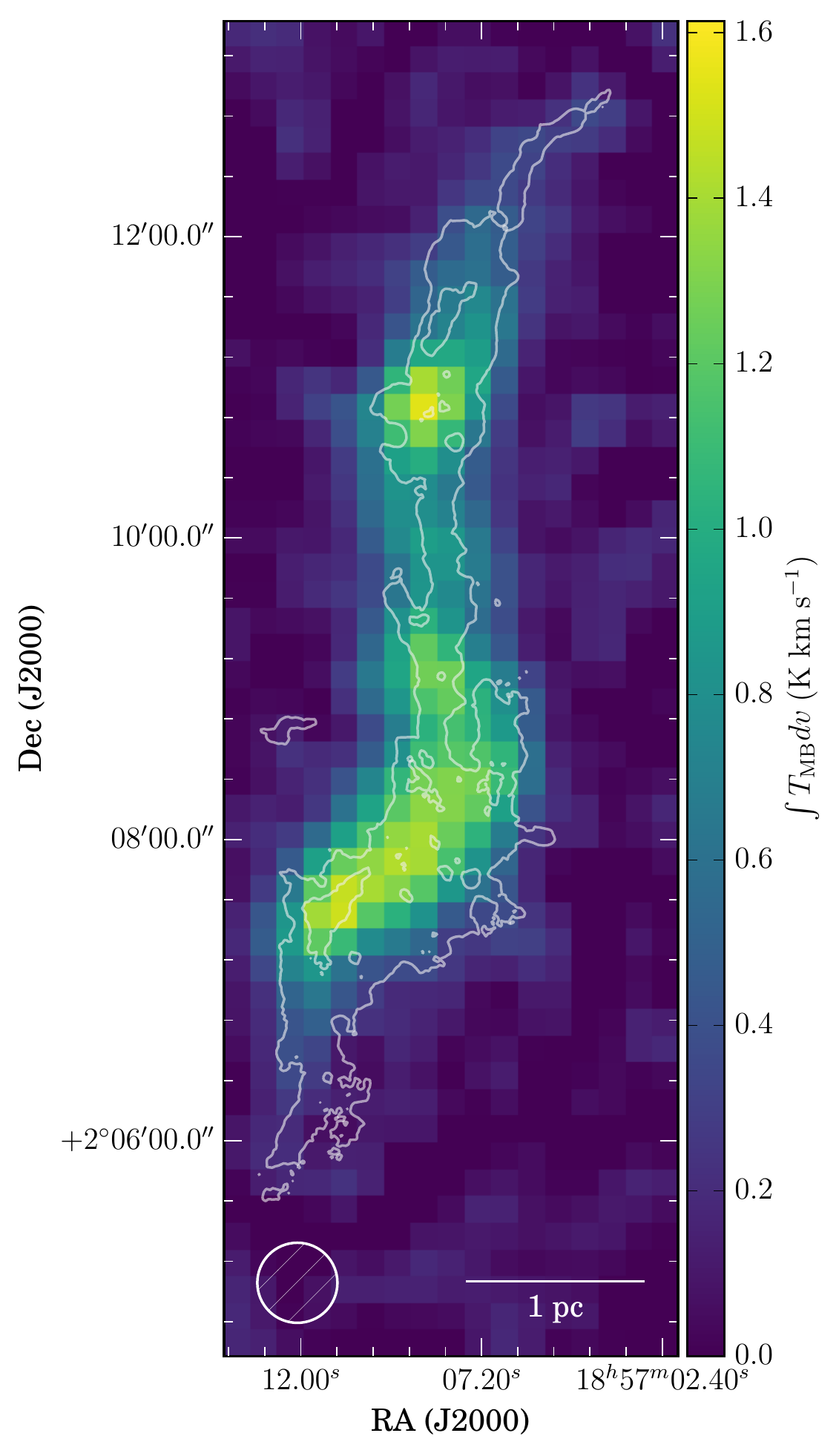}{(b)}{white}{0.33}
    \subfigcolorsizelabel{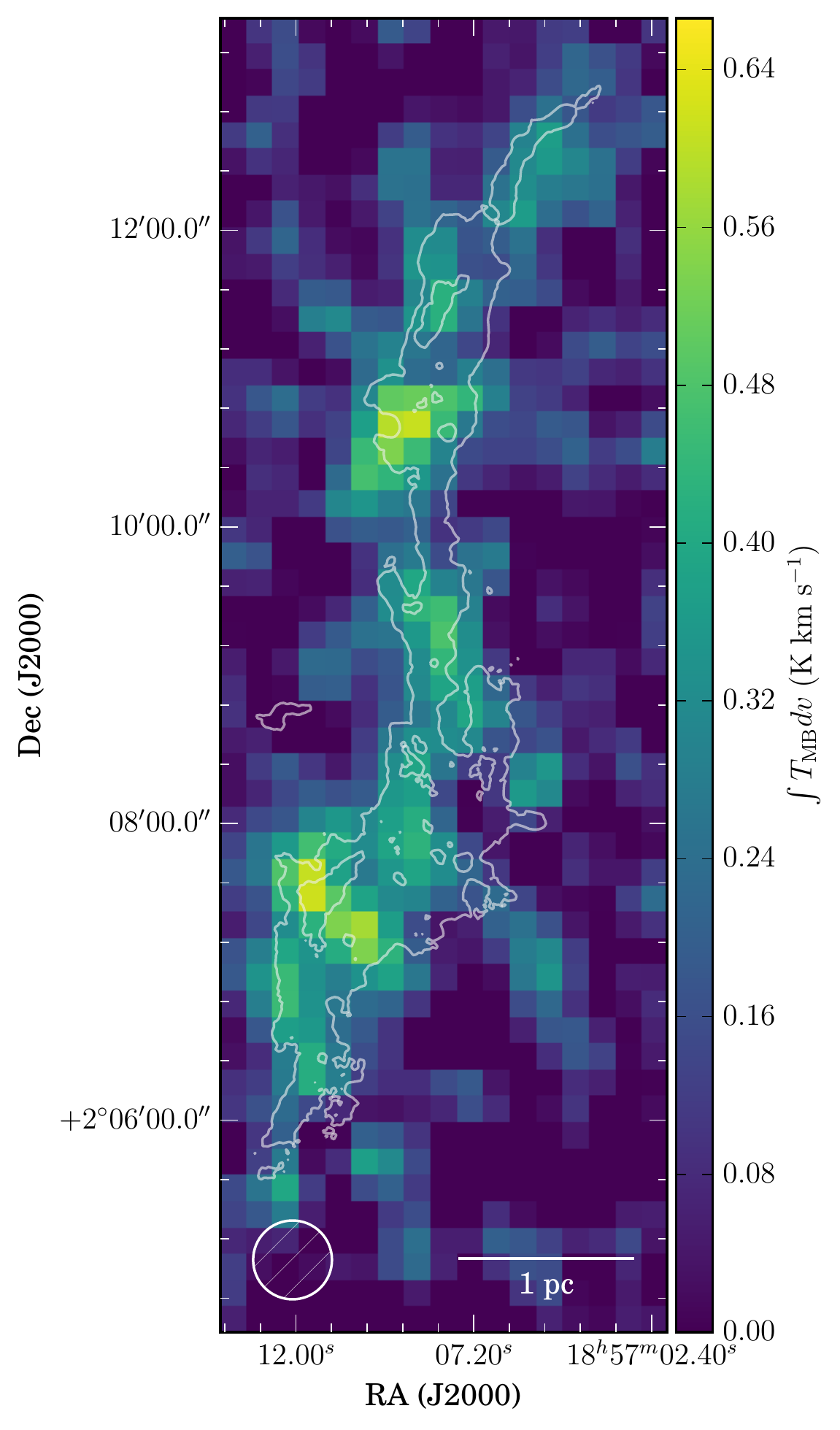}{(c)}{white}{0.33}
    \caption{Data products from GBT obervations of G035.39. Integrated intensities of \nhhh inversion transitions (1, 1) and (2, 2), \textit{(a)} and \textit{(b)}, respectively; and integrated intensity of the rotational CCS (2-1) line \textit{(c)}. \nhhh (2, 2) and CCS lines were integrated between 43 and 47 \kms to highlight the emission features, while the \nhhh (1, 1) line was integrated between 20 and 70 \kms. Extinction contours from \cite{kainulainen+tan2013} at $\mathrm{A_V} = 25$ and 65 mag are shown in white.}
    \label{fig:integrated}
\end{figure*}

\begin{figure*}[t!]
    \centering
    \includegraphics[width=1.0\textwidth]{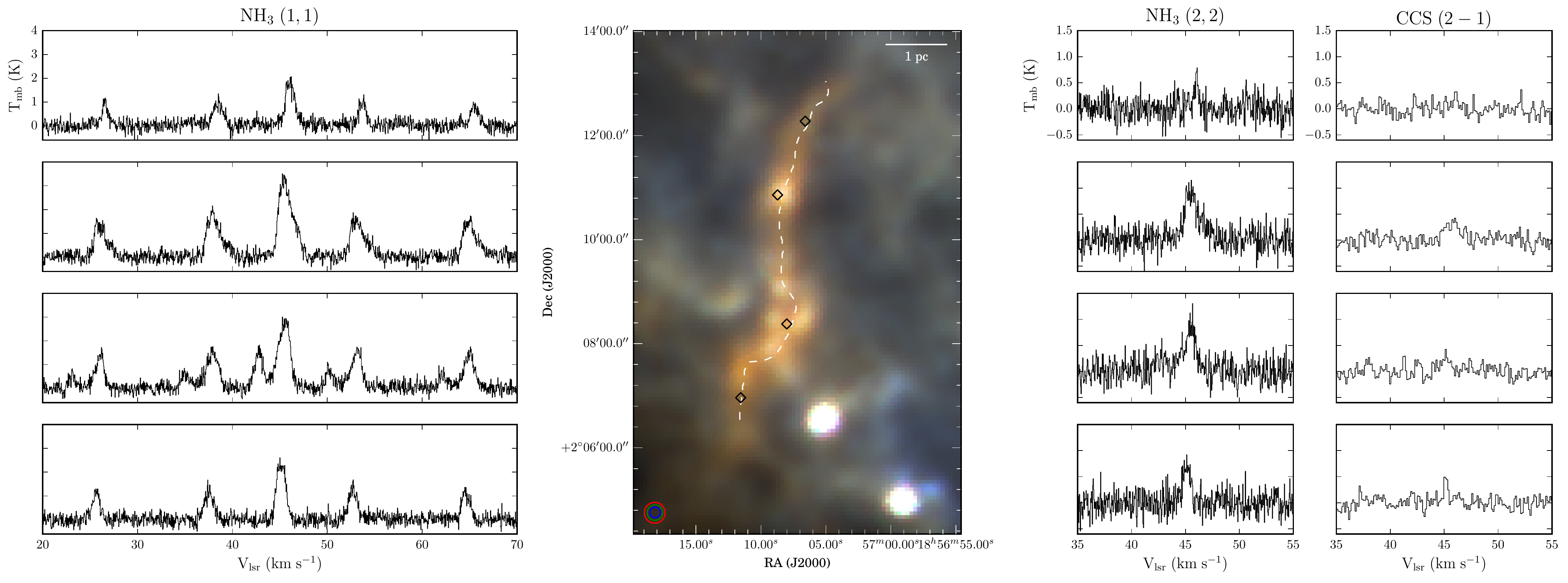}
    \caption{A composite Hi-GAL color image of the infrared dark cloud. Red, green, and blue channels correspond to 350, 250, and 160\micron, respectively. To the left of the \herschel map, four \nhhh (1, 1) spectra are shown, and four selected \nhhh (2, 2) and CCS (2-1) lines are displayed on the right side of the central figure. The CCS spectra are smoothed to 0.12 \kms. Each row of spectra is arranged to match the black diamond markers shown on the FIR map, selected to represent typical spectra along the filament. The filament ridge, as defined in \S\ref{sec:gbtobs}, is marked with a dashed white line.}
    \label{fig:higalspec}
\end{figure*}

The integrated intensity maps for ammonia and CCS transitions observed with the GBT are presented in Fig. \ref{fig:integrated}. Not accounting for the bright \nhhh satellite lines, the majority of the molecular emission detected from the cloud is situated between 43 and 46 \kms. A prominent secondary component is present in the southern region of the cloud and, furthermore, strong line asymmetries indicate the presence of additional line-of-sight components (see Fig. \ref{fig:higalspec} for example spectra).

Given the low signal-to-noise ratio of the CCS (2-1) line, its kinematics can not be easily constrained, and the transition is not discussed throughout this work. We note that the southern peak of the integrated intensity of the CCS coincides with a peak of the mid-infrared extinction map (Fig. \ref{fig:integrated}c), and has no associated 24- and 70\micron point sources. As the carbon-chain molecules are known to trace regions of early-stage chemistry \citep{suzuki+1992},
the CCS peak may be indicative of dense gas in an earlier evolutionary stage than the rest of the cores in the IRDC.
 More dedicated studies will be carried out toward this region in the future.

As the large-scale structure of the cloud's main velocity component manifests itself as a prominent, continuous filament, we can probe how its physical properties vary along its length.
We set the line segment defining the filament from the ammonia intensity profile. The maximal value of the \nhhh (1,1) integrated intensity along right ascension is calculated for each value of declination. The longest continuous segment is then taken from the obtained set of coordinate points, with a continuity condition that the neighboring points must be in the same GBT beam. This approach, adequately simplistic for our task of probing pc-scale structure, recovers the intensity ridge of G035.39 without the need to resort to more sophisticated ridge detection methods. The resulting filament profile is marked on Fig. \ref{fig:overview}.

In the far-infrared \herschel photometric maps, G035.39 manifests itself as a typical infrared dark cloud, characterized by its emission at longer wavelengths. It appears as an extinction feature in the 70\micron band of PACS, and as an emission feature from 160\micron onwards. The central panel on Fig. \ref{fig:higalspec} displays a color-composite image of the IRDC, with 350, 250, and 160\micron emission used for red, green, and blue channels, respectively.
As both ammonia and dust continuum trace the dense gas that constitutes the bulk of the cloud, the far-infrared morphology does not show any significant deviations from the \nhhh-defined filament ridge. A clearly visible substructure in PACS 160\um and SPIRE 250\um maps can be attributed to those \herschel bands having angular resolutions (12\arcsec~and 18\arcsec, respectively) different from our \nhhh observations (32\arcsec).
The properties derived from the \herschel maps were then regridded onto the grid given by our GBT observations.

\subsection{Ammonia line fitting}
\label{sec:ammonia}

The ammonia molecule has been proven to be an invaluable tool in probing physical conditions of moderately dense molecular regions \citep[e.g.][]{myers+benson1983, rosolowsky+2008-nh3, pillai+2006-nh3, pineda+2010, wang+2012, wang+2014}, and the ratio of its collisionally populated metastable $(J,K) = (1,1)$ and $(2,2)$ inversion states can be used to derive the rotational gas temperature \Trot \citep{ho+townes1983}.
This temperature is commonly used as a proxy value for the kinetic temperature of the medium, derived through balancing the rates of radiative and collisional transitions \citep{walmsley+ungerechts1983, tafalla+2004, swift+2005}.

\begin{figure*}
    \centering
    \subfigfivecolorlabel{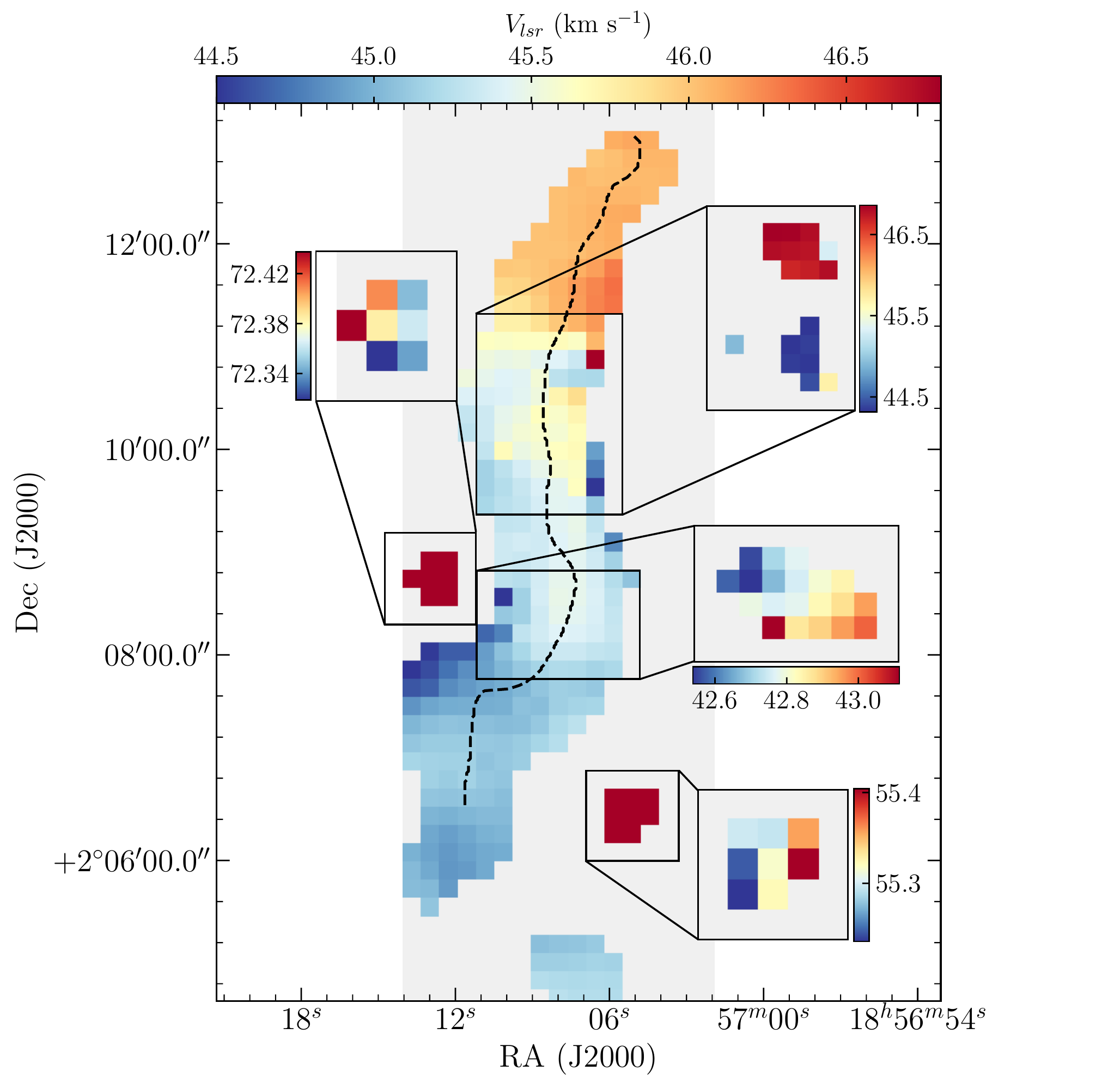}{(a)}{black}
    \subfigfivecolorlabel{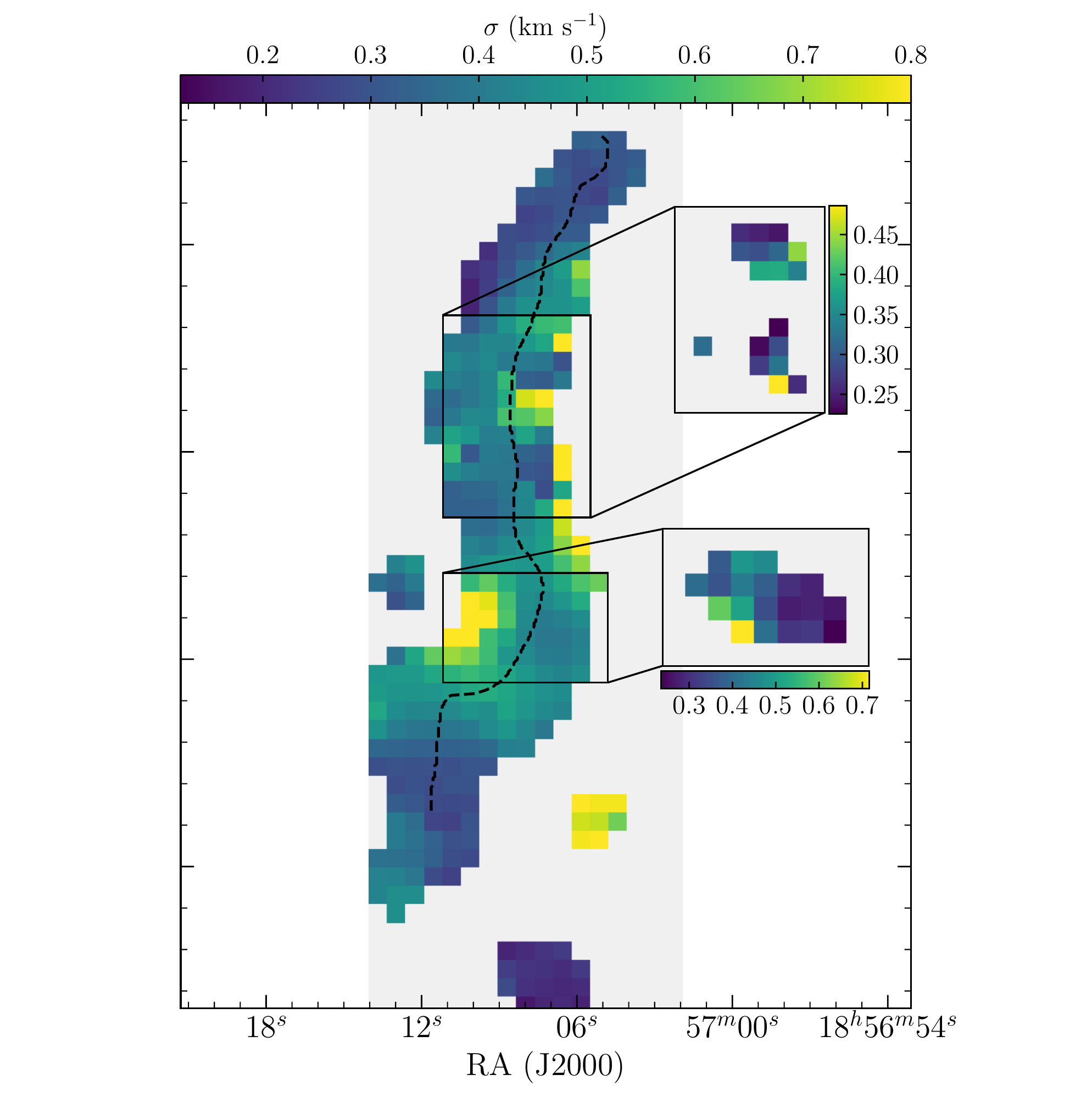}{(b)}{black}
    \caption{a) Fitted velocities relative to line rest frequency. Overlaid on the main velocity component, additional components, enclosed in the black contours, are shown in the cut-outs. The ridge of G035.39, as defined in \S\ref{sec:gbtobs}, is marked as a dashed black line. b) Same as (a), but for the fitted velocity dispersions.}
    \label{fig:kinematics2d}
\end{figure*}

In the Rayleigh–Jeans approximation, an observed spectral profile of the two inversion lines along with their hyperfine components can be described as a sum over their hyperfines, assuming uniform excitation conditions for all the hyperfine components of the lines
and a unity beam filling factor \citep[e.g.][]{stahler+palla}:
\begin{equation} \label{eq:line_profile}
T_{\mathrm{MB}}(v) = \sum_i\left\{1 - \mathrm{exp}\left[-r_i~\tau_0~\mathrm{exp}(-\frac{(\voff-v_i)^2}{2\sigma_v^2})\right]\right\}(\Tex - \Tbg),
\end{equation}
where \Tex describes the population ratios of the inversion transition parity levels, the background radiation temperature \Tbg is taken to be that of the cosmic microwave background (2.7312 K), $r_i$ are normalized relative intensities of the hyperfine components, $v_i$ is the velocity centroids of the hyperfine components, $\sigma_v$ is the velocity dispersion, and $\tau_0$ the main group opacity of the corresponding inversion line. The latter can be used to derive the column density
of the (1,1) rotational state $N_{(1,1)}$ \citep{rosolowsky+2008-nh3}:
\begin{equation} \label{eq:N11}
N(1,1) = \frac{8 \pi k \nu_0^2}{h c^3} \frac{1}{A_{1,1}} \sqrt{2\pi}\sigma_\nu (T_\mathrm{ex}-T_\mathrm{bg})\tau_0,
\end{equation}
and can then be scaled to a total ammonia column density \Ntot via a partition function,
assuming that only the metastable population levels are populated \citep{rosolowsky+2008-nh3}.
The spectral profile modelling and line fitting are done following the method presented in 
Friesen, Pineda et al. \citeyearpar{friesen+pineda+2017},
where the \nhhh (1,1) and (2,2) spectra are simultaneously modelled using Eq. \ref{eq:line_profile} within the \texttt{pyspeckit} \citep{pyspeckit} Python package. The spectral profile of both inversion lines are fitted considering the following free parameters: kinetic temperature \Tkin, excitation temperature \Tex, ammonia column density \Ntot, velocity offset with respect to line rest frequency \voff, and velocity dispersion \sig.

As mentioned in \S\ref{sec:dataoverview}, multiple line-of-sight components are present in the ammonia spectra. Their interchangeable strength and unknown a priori locations make conventional line-fitting techniques difficult to apply.
In particular, the iterative nonlinear least-squares algorithm \citep{levenberg1944, marquardt1963} used in \texttt{pyspeckit} is prone to failures to reach global convergence in presence of multiple local minima.
To ensure convergence of the algorithm on the global minima, we perform parameter space gridding 
to select initial guesses for the fitting routine. As the parameters that vary the most among our spectra are the line centroids and their peak amplitudes, we search for the optimal starting conditions by varying \Tex and \voff.

By inspecting the \nhhh (1,1) spectra for line peaks and line asymmetries, we have set the velocity intervals at which the line centroids could reside. The six intervals selected are centered at 42.9, 44.6, 45.5, 47.0, 55.2, and 72.2 \kms. The velocity components that were found to overlap spatially were each split into 10 values, corresponding to a velocity steps of $\leq 0.1$ \kms. Additionally, we consider a range of 10 \Tex values set up to cover the amplitude range of our \nhhh (1, 1) data. The resulting velocity and line brightness ranges for components pairs that were found to overlap were then permuted together to yield over 50k modeled spectra. Each modeled spectral profile was subsequently checked against all the pixels in our data, and the models with lowest total squared residuals were used as a starting point for the nonlinear least squares routine.
The source code for the initial guess selection method described above is freely available online\footnote{\url{https://github.com/vlas-sokolov/multicube}}.

For every pixel of the spectral cube obtained from \S\ref{sec:gbtobs} we perform a multiple component nonlinear least-squares fit for one and two velocity components. To decide on a number of components present in a spectrum, one needs to rule out overfitting. As a direct minimization of the square of residuals would always prefer a more complex model, we limit the multiple component fits, requiring the best-fit solution to have a signal-to-noise ratio of at least three in all components and for those components to have the peak separation larger than the line widths of the two components. Should any of these criteria not be met, a simpler model is preferred - one velocity component for a failed double peak fit, or a spectrum is masked altogether if a single peak fit fails.

\subsection{Parsec-scale kinematics of the IRDC}
\label{sec:kinematics}

The structure of G035.39 can be seen as a combination of a main velocity component and additional velocity components along the line of sight. We refer the coherent structure at 44-47 \kms as the main velocity component. For spectra with two line components identified, components that are brighter than their counterparts are considered to belong to the main group. In G035.39, this choice results into a spatially coherent velocity field traced by the main component.

Figure \ref{fig:kinematics2d} illustrates the kinematics of the main velocity component, overlaid with additional velocity components identified. Of these additional components, the ones found in G035.39-N (the 44 and 46 \kms components) coincide with the network of filaments from \cite{henshaw+2014}.
Two more velocity components are identified in both ammonia transitions from the spectral cube inspection, but at $55$ \kms and $72$ \kms are unlikely to be related to the main body of the IRDC.
Additionally, a strong (up to peak $T_{\mathrm{MB}} = 1.5~\mathrm{K}$) velocity component, well-separated from the main cloud component by $\sim$2 \kms, is present at the location of the active star formation in the southern part of the IRDC. This is consistent with the location and \vlsr of Filament 1 from \cite{jimenez-serra+2014} at its southernmost point.

\begin{figure}
    \centering
    \subfigcolorrightlabel{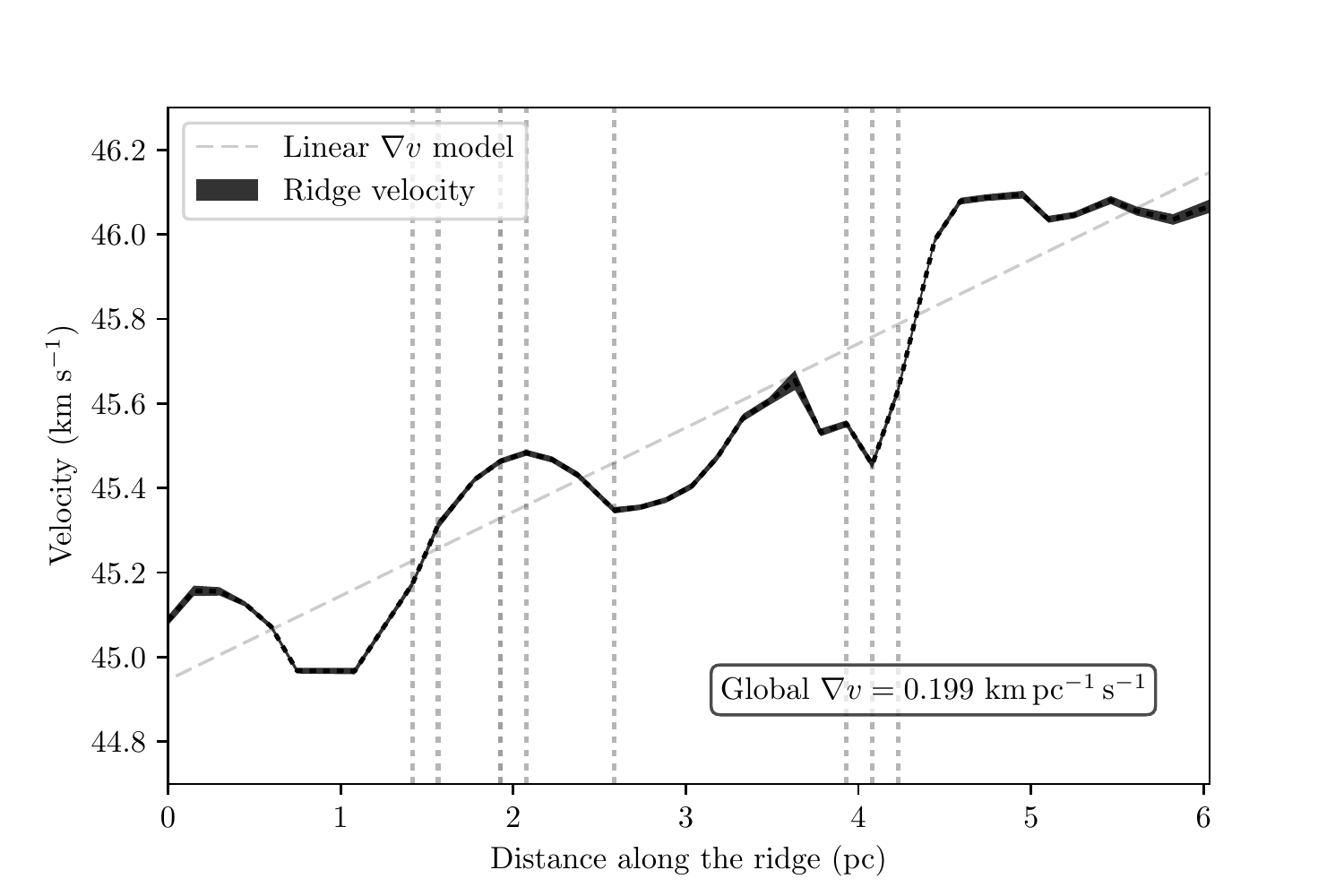}{(a)}{black}
    \subfigcolorrightlabel{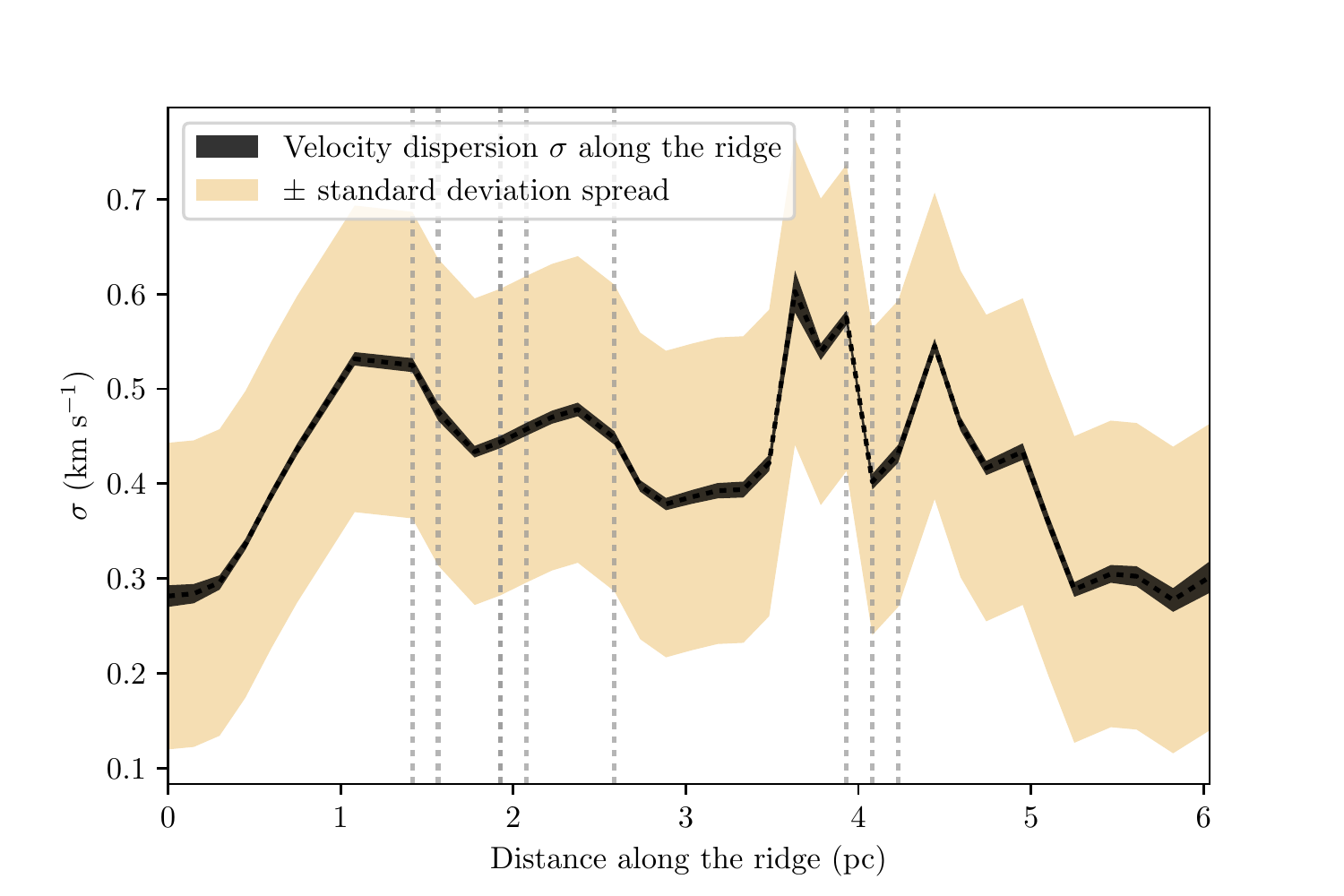}{(b)}{black}
    \caption{(a): The radial velocity profile of the main component along the G035.39 filament, starting at its southernmost point. The solid black line shows the radial velocities fit at the intensity ridge of the IRDC. The width of solid black line represents the fitting uncertainty at each point. The vertical dotted lines mark the projected locations of the massive protostellar cores from \cite{nguyen_luong+2011}. (b): same as (a), but for the velocity dispersion profile along the IRDC ridge. The yellow shaded region denotes the spread area given by two standard deviations of all $\sigma$ values in the main velocity component.}
    \label{fig:velgrad}
\end{figure}

The velocity of the main ammonia component
gradually changes from red- to blue-shifted in the southward direction.
Figure \ref{fig:velgrad} illustrates this change, showing the velocity centroid and velocity dispersion profiles along the IRDC intensity ridge.
The non-thermal velocity dispersions of ammonia \citep{myers1983} have a large dynamic range,
sometimes going as high as 1.2 \kms at the edges of the map,
but generally staying within the interquartile range between 0.38 and 0.52 \kms (Fig. \ref{fig:velgrad}b).
These high values, implying non-thermal motions dominating the line width, are above those found in low-mass cores and are in the upper range of typical non-thermal components of massive cores \citep{caselli+myers1995}.
For the $\mathrm{H_2}$ sound speed derived from the fitted ammonia kinetic temperatures and assuming a mean mass per particle of 2.33 u, the average Mach number across the IRDC is $\mathcal{M} = 2.14$, consistent with previous studies of G035.39 that find gas motions in the cloud to be supersonic \citep{henshaw+2014, jimenez-serra+2014}.

The gradual change of the line centroid towards the southern portion of the IRDC can be quantified in terms of a velocity gradient.
Previous studies of G035.39 have discussed the global gas motions on various scales.
\cite{henshaw+2014} attribute radial velocity irregularity towards the northern part of G035.39-N found by \cite{henshaw+2013} to the unresolved substructure, and find the global gradients in identified filaments to be smaller than 0.7 \kmspc.
\cite{jimenez-serra+2014} find global, north-south velocity gradients of ${\sim}0.4-0.8$ \kmspc along three CO filaments in G035.39.
As our GBT data covers the full extent of the cloud, we are able to constrain the global velocity gradient along the whole IRDC. A least squares fit to the ridge velocity profile, weighed by uncertainties in centroid velocities, results in a line-of-sight velocity gradient of $\nabla v \sim 0.2$ \kmspc along the filament's 6 pc length (Fig. \ref{fig:velgrad}a).
This value is in good agreement with the average velocity gradients reported for larger scale filament and GMC structures \citep[e.g.,][]{hernandez+tan2015, wang+2016}.

\begin{figure}
    \centering
    \includegraphics[width=0.45\textwidth]{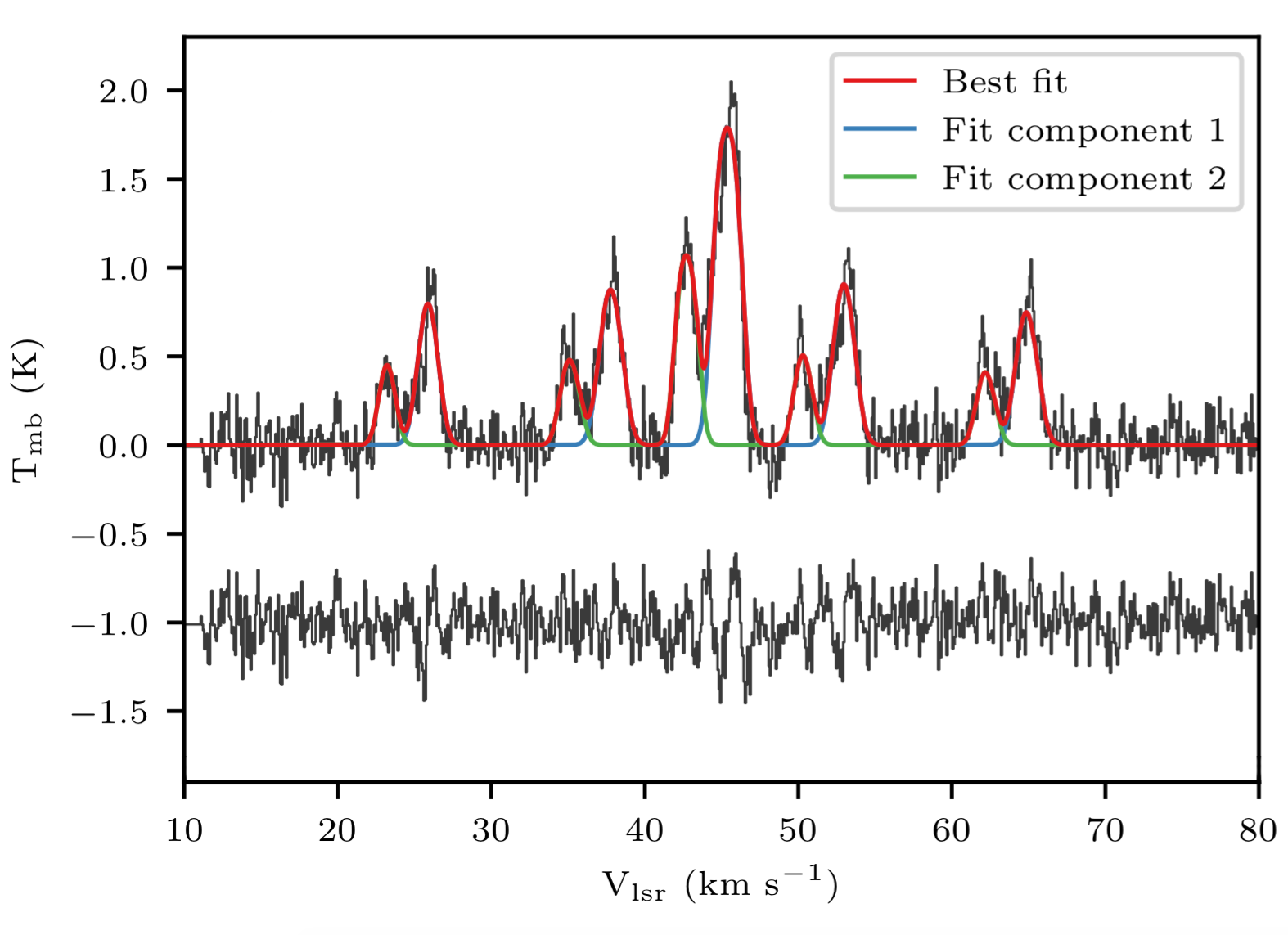}
    \caption{\nhhh (1,1) spectrum towards $\mathrm{\alpha(J2000) = 18h57m10.1s}$ and $\mathrm{\delta(J2000) = +2\arcdeg08\arcmin23\arcsec}$, overlaid with the best two-component fit model. A significant residual, revealing the presence of the unconstrained third component, is plotted alongside the spectrum, offset by 1 K. The spectrum was smoothed to 0.08 \kms for visual clarity.}
    \label{fig:nh3-x2}
\end{figure}

While the global velocity gradient following the ridge of G035.39 is well described on scales larger than one parsec, smaller scale, oscillatory-like deviations from the fitted linear relation are present.
Additionally, hints of a localized third component, towards the southern 42.6 \kms feature, are manifested as a broader velocity dispersion at the edge of G035.39 (see Fig. \ref{fig:nh3-x2} for an example fit with irregular residual). As it is mostly blended with the main component, constraining its properties is difficult due to the limited angular resolution of the GBT data.
Ammonia emission is often highly substructured in star-forming regions, and can manifest itself as filamentary emission down to ${\sim}5000$ AU scales \citep{pineda+2011, pineda+2015}.
The exact nature of the velocity substructure in the southern part of G035.39 can be seen as either stemming from the coherent gas motions around the dense cores \citep[e.g., ][]{hacar+tafalla2011, zhang+2015, gritschneder+2017}, or as a picture reminiscent of the nearby low-mass star forming Taurus complex, where previously unresolved velocity-coherent filaments are found to be bundled together in a larger structure \citep{hacar+2013, tafalla+hacar2015}.
The velocity irregularities along the cloud may result from the sub-pc substructure emission (similar to the already resolved one in the G035.39-N region by \cite{henshaw+2014}), being smoothed by the GBT beam.
The higher angular resolution analysis of the gas kinematics along the entire IRDC will be addressed in a future study.

\subsection{Gas temperature}
\label{sec:gas}

While the profile of the \nhhh (1, 1) line can be constrained accurately in the pixels with significant emission, the (2, 2) line is considerably weaker in the low temperature regime. Because of this, we impose a more stringent constraint on the physical parameters that depend primarily on the (2, 2) inversion transition, namely kinetic temperature and total ammonia column density. For these, we only select spectra that have $> 3 \sigma_I$ detection in the integrated intensity. We take the expression for integrated intensity uncertainty, $\sigma_I = \sqrt{N} \sigma_{T_{\mathrm{MB}}} \Delta v_c$, from \cite{mangum+shirley2015}, where $N$ is the number of channels with nonzero emission in a modeled spectrum (arbitrarily taken to be $10^{-5}$ K), $\sigma_{T_{\mathrm{MB}}}$ is the corresponding $\mathrm{T_{MB}}$ uncertainty, and $\Delta v_c$ is the channel width in the GBT spectra.

\begin{figure}
    \centering
    \makebox[\textwidth][l]{\includegraphics[width=0.42\textwidth]{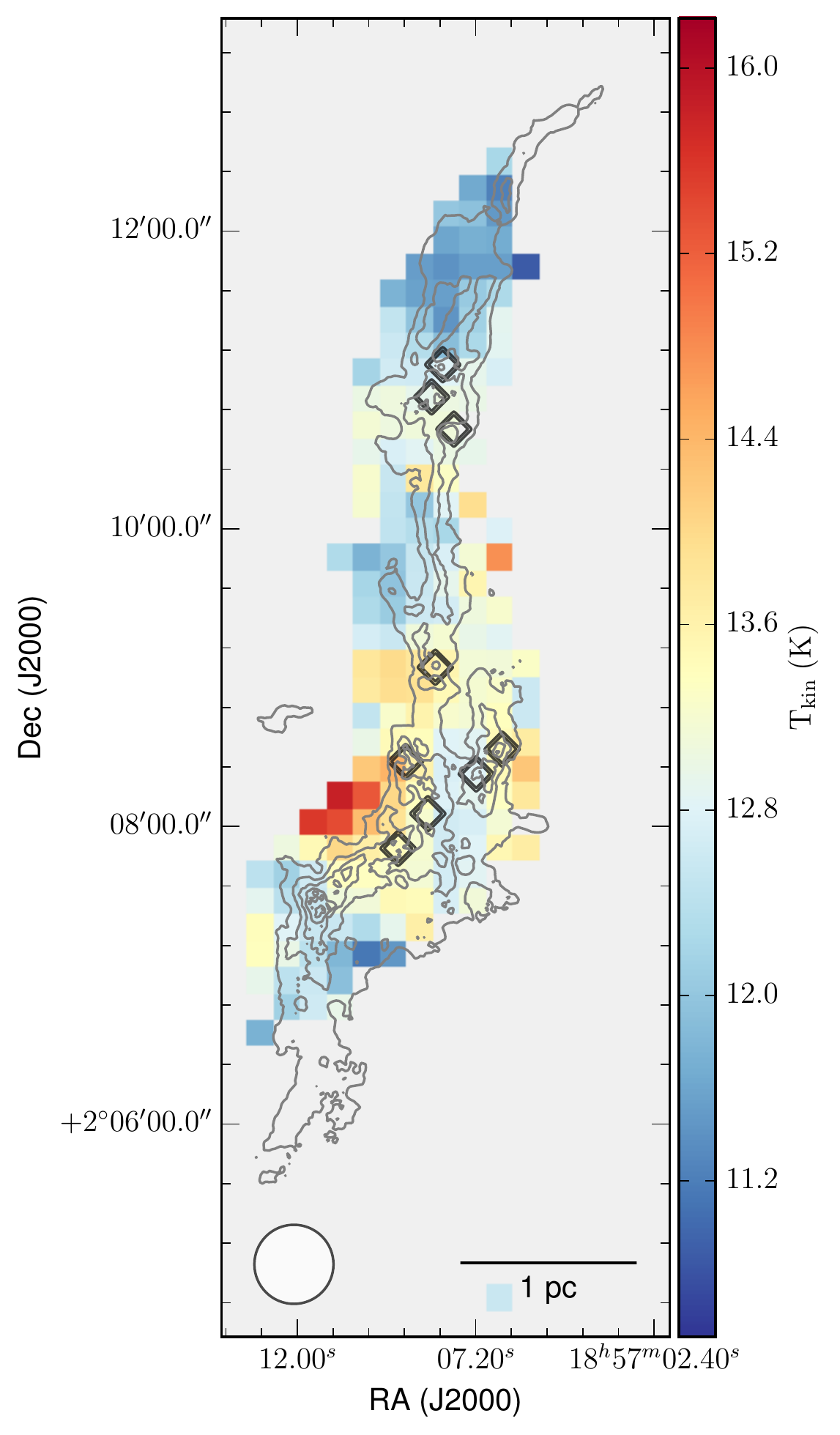}}
    \caption{Kinetic temperature map derived from the main velocity component of \nhhh. The contour lines and diamond markers are the same as on Fig. \ref{fig:overview}.}
    \label{fig:temp_main}
\end{figure}

The gas kinetic temperature of the main component of G035.39 is presented in Figure \ref{fig:temp_main}. The derived values vary across the body of the dark cloud from about 11 K to 15 K. The cold gas temperature range is typical of other IRDCs \citep{pillai+2006-nh3, ragan+2012, chira+2013, wang+2012, wang+2014} as well as of relatively nearby low-mass starless cores \citep{rosolowsky+2008-nh3, friesen+2009, forbrich+2014}.

Due to a more stringent masking of the temperatures derived from our multi-component fit, only a handful of spectra have their kinetic temperature constrained for weaker secondary components. As most of the derived values of the gas kinetic temperature belong to the main component, the discussion on the temperature of the weaker components is thus effectively restricted to those few detections. In the star-forming part of G035.39-N, towards the northern group of the star-forming cores, we successfully measure the gas temperature for two components along the line of sight to be between 12.9 and 13.3 K.

The region coincident with a bright infrared source south-east of the main filament (55 \kms cut-out on Fig. \ref{fig:kinematics2d}) is consistently hotter than the bulk of the IRDC. The derived temperatures for the gas associated with the source range from 17.5 to 20.9 K, suggesting that the gas is internally heated. \cite{nguyen_luong+2011} identify the source as an infrared-bright protostellar MDC, and derive its dust temperature to be $26 \pm 6$ K. As the ammonia emission for this component is peaked at 55 \kms, it appears to be physically unrelated to G035.39. Similarly, for a spectral line detected at 72 \kms (Fig. \ref{fig:kinematics2d}), which appears as a starless infrared extinction feature east of the filament, we report a kinetic temperature of $16.5 \pm 1.3$ K. Although defined as part of the IRDC extent in \cite{nguyen_luong+2011} based on the common column density contour, we suggest that the feature does not form a coherent structure with G035.39.

\subsection{Dust temperature}
\label{sec:dust}

\begin{figure*}
    \subfigcolorlabel{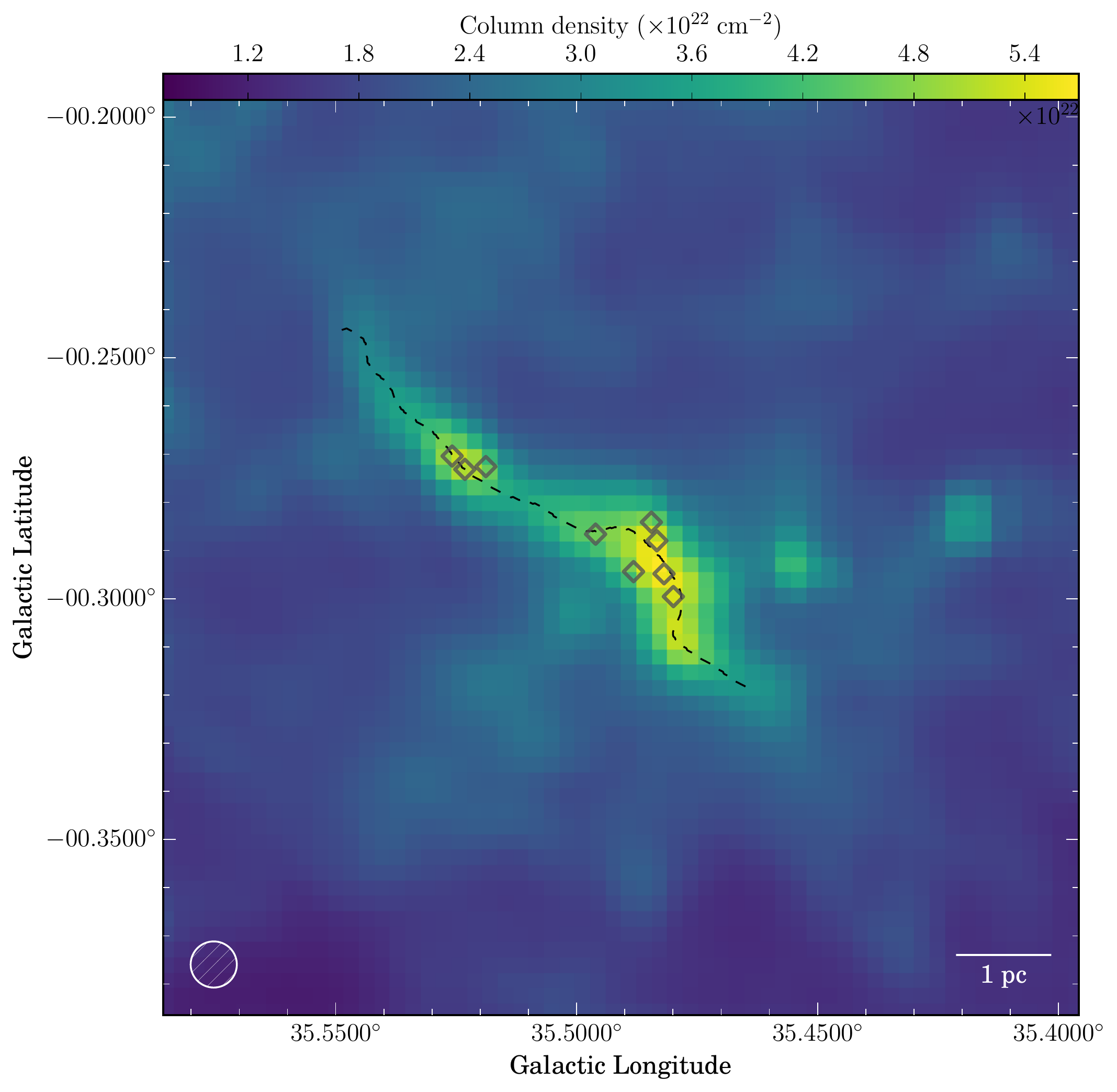}{(a)}{white}
    \subfigcolorlabel{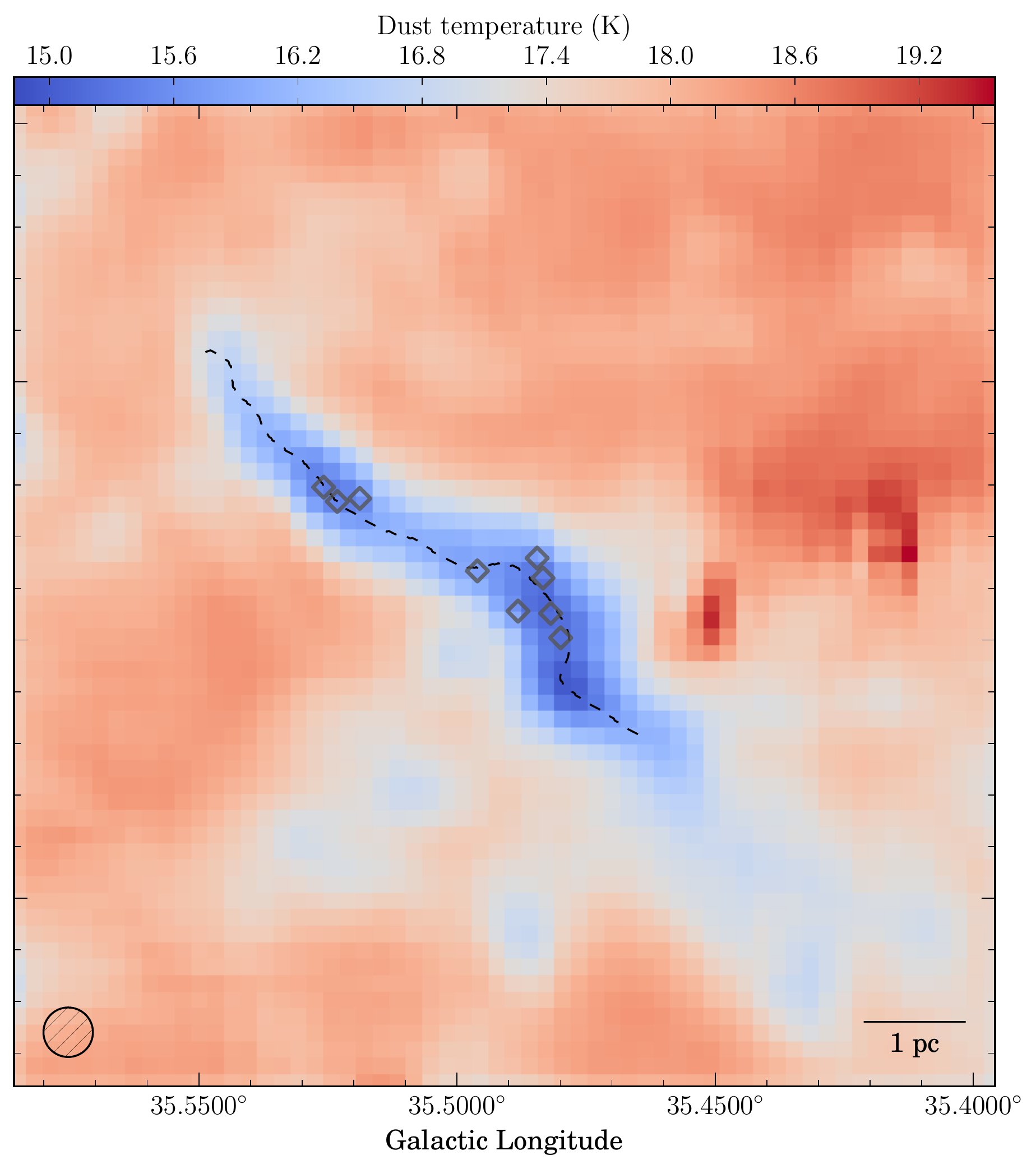}{(b)}{black}
    \caption{$\mathrm{H_2}$ column density (a) and dust temperature (b) maps derived from \herschel data for G035.39. A 1 pc length scale and a combined \herschel beam size are shown for each map. The ammonia intensity ridge and the positions of \herschel MDCs are shown as a black dashed line and grey diamond markers, respectively.}
    \label{fig:higal_TN}
\end{figure*}

The intensity of an optically thin source of temperature $T$ is given by the radiative transfer equation \citep[e.g.,][]{rybicki_lightman}, which can be approximated by
\begin{equation} \label{eq:intensity}
I_{\nu} = B_{\nu}(T)\times(1 - e^{-\tau_{\nu}}) \approx B_{\nu}(T) \tau_{\nu},
\end{equation}
where $B_{\nu}(T)$ is the Planck blackbody function. In the expression above, a frequency dependent opacity $\tau_{\nu}$ can be shown to be related to the gas column density $N_{\mathrm{H_2}}$ in the following way:
\begin{equation} \label{eq:opacity}
\tau_{\nu} = \int \kappa_{\nu}\rho\mathrm{d}s = N_{\mathrm{H_2}}\mu_{\mathrm{H_2}}m_{\mathrm{H}}\kappa_{\nu},
\end{equation}
where $\mu_{\mathrm{H_2}} = 2.8$ is the molecular weight per hydrogen molecule \citep{kauffmann+2008}, $\kappa_{\nu}$ is the dust opacity, and $m_{\mathrm{H}}$ is the mass of a hydrogen atom.
For consistency with \cite{nguyen_luong+2011}, we have adopted the dust opacity law of $\kappa_{\nu} \propto \nu^{\beta}$, assuming a gas-to-dust mass ratio of 100 from \cite{hildebrand1983}. As the optically thin approximation in Eq. \ref{eq:intensity} does not necessarily hold for PACS bands at high column densities, we discuss its limitations for the PACS data. While the 160\um PACS band is expected to be within the optically thin approximation well into the high density regime ($\tau_{160} = 0.16$ for $N_{\mathrm{H_{2}}} = 10^{23}~\mathrm{cm^{-2}}$), the 70\um maps may trace extinction of dust rather than its emission (G035.39 appears as extinction feature in 70\micron), rendering the optically thin dust emission assumption invalid \citep[e.g.,][]{battersby+2011}.

Hi-GAL PACS and SPIRE maps containing G035.39 were jointly convolved to a common resolution of 34.5\asec, corresponding to the beam size of the 500\micron SPIRE band. The maps were then regridded to 11.5\asec pixel size to form spectral cubes.
The resulting cube was fit pixel-by-pixel with a single temperature gray-body model described above, assuming a fixed value of $\beta = 2$ \citep[e.g.][]{stutz+2010, nguyen_luong+2011, elia+2013, lombardi+2014}.
Fixing the value of $\beta$ reduces the number of free parameters for an otherwise degenerate model \citep[e.g.,][]{planck2011, kelly+2012}. While implicit assumptions on the spectral opacity index are made in such a way, it enables an estimation of the resulting parameter uncertainties by minimizing the errors stemming from the model degeneracy.
In the analysis above, we have considered 30\% uncertainty on the flux levels for consistency with \cite{nguyen_luong+2011}.

Column density and effective dust temperature maps, derived for the \textit{Herschel} regions from 160, 250, 350 and 500\micron \textit{Herschel} bands, are presented in Figure \ref{fig:higal_TN}.
The column density and dust temperature values show anti-correlation, typical for externally irradiated clouds (\citeauthor{evans+2001} \citeyear{evans+2001}, see also \citeauthor{planck2011} \citeyear{planck2011}),
and IRDCs \citep[e.g.,][Appendix]{wang+2015}.
The overall morphology of the density and temperature structure in the derived parameter maps is similar to that of \cite{nguyen_luong+2011}.

\subsection{Molecular abundances}
\label{sec:abunances}

\begin{figure}[t!]
    \centering
    \includegraphics[width=0.4\textwidth]{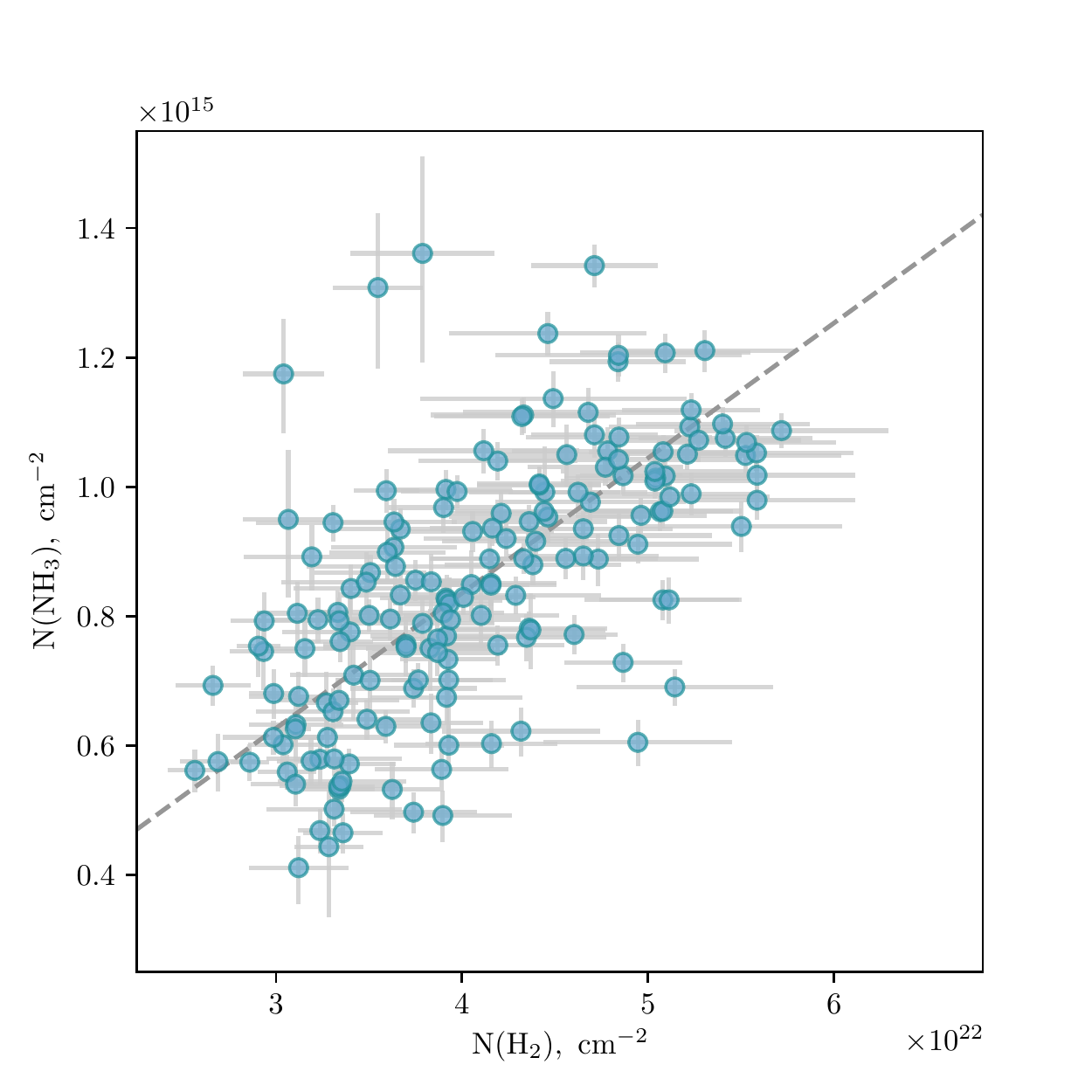}
    \caption{\herschel-derived $\mathrm{H_2}$ gas column density plotted against fitted \nhhh column density. The dashed straight line passes through both the mean ammonia abundance in the IRDC and through the zero column density point.}
    \label{fig:NvN}
\end{figure}

The column densities from \herschel and those of ammonia correlate weakly ($r=0.64$, Fig. \ref{fig:NvN}), and show no evidence for ammonia depletion, consistent with previous work toward nearby starless cores \citep{tafalla+2002}. By taking the ratio of the two we derive the ammonia abundance in G035.39, 
$X(\mathrm{NH_3}) = [N(\mathrm{NH_3})/N(\mathrm{H_2})]$. 
The mean value of the total \nhhh abundance is ${\sim}2.1 \times 10^{-8}$, corresponding to the para-\nhhh abundances of ${\sim}1.0 \times 10^{-8}$ assuming the ortho- to para-\nhhh ratio of one. The range of abundances is comparable to that found towards the centers of low-mass cores (\citealt{tafalla+2004}; \citealt{crapsi+2007}; Friesen, Pineda et al. \citeyear{friesen+pineda+2017}), and is in good agreement to values measured towards the low-mass star-forming cluster Serpens South \citep{friesen+2016}. When compared to other IRDCs, our GBT data on G035.39 point to lower abundances - \cite{ragan+2011} derive \nhhh abundances of a few times $10^{-7}$. However, \citeauthor{ragan+2011} observe their IRDC sample with a much smaller, VLA, beam sizes, so allowing for a small beam filling factor of our observations may mitigate the discrepancy. Indeed, for angular resolution of 40\asec and a median sample distance of 2.9 kpc, \cite{pillai+2006-nh3} report $X(\mathrm{NH_3})$ to be between $7 \times 10^{-9}$ and $10^{-7}$, consistent with our findings.

\section{Discussion}
\label{sec:discussion}

\subsection{A comparison of dust and gas temperatures}
\label{sec:temperature}

Current understanding of heating and cooling processes within dense molecular clouds requires close coupling between gas and dust at number densities above ${\sim}10^{5}~\mathrm{cm^{-3}}$ \citep{goldsmith2001}.
However, no correlation of dust and gas temperatures is found in G035.39 (Pearson's $r = -0.16$), despite the uncertainties on the temperatures being significantly smaller than the temperature dynamical range.
Moreover, the dust temperatures derived from \herschel are consistently higher than the ammonia based gas temperatures, suggesting that the two sets of observations are sensitive to different environments.
Dust temperatures higher than the gas temperatures have been reported before. \cite{forbrich+2014} find dust temperatures to be $2-3$ K warmer than the gas temperatures towards starless cores in the Pipe nebula. More recently, \cite{friesen+2016} report their ammonia-derived kinetic gas temperatures to be consistently lower than their \herschel counterparts. The difference in the two temperature tracers can be attributed to the line-of-sight mixing of warm foreground dust emission into the far-infrared SED, thus raising the effective line-of-sight average dust temperature. Ammonia gas temperature, on the other hand, is expected to trace the dense inner region of the filament only.

\begin{figure}
    \centering
    \subfigcolorrightlabelnew{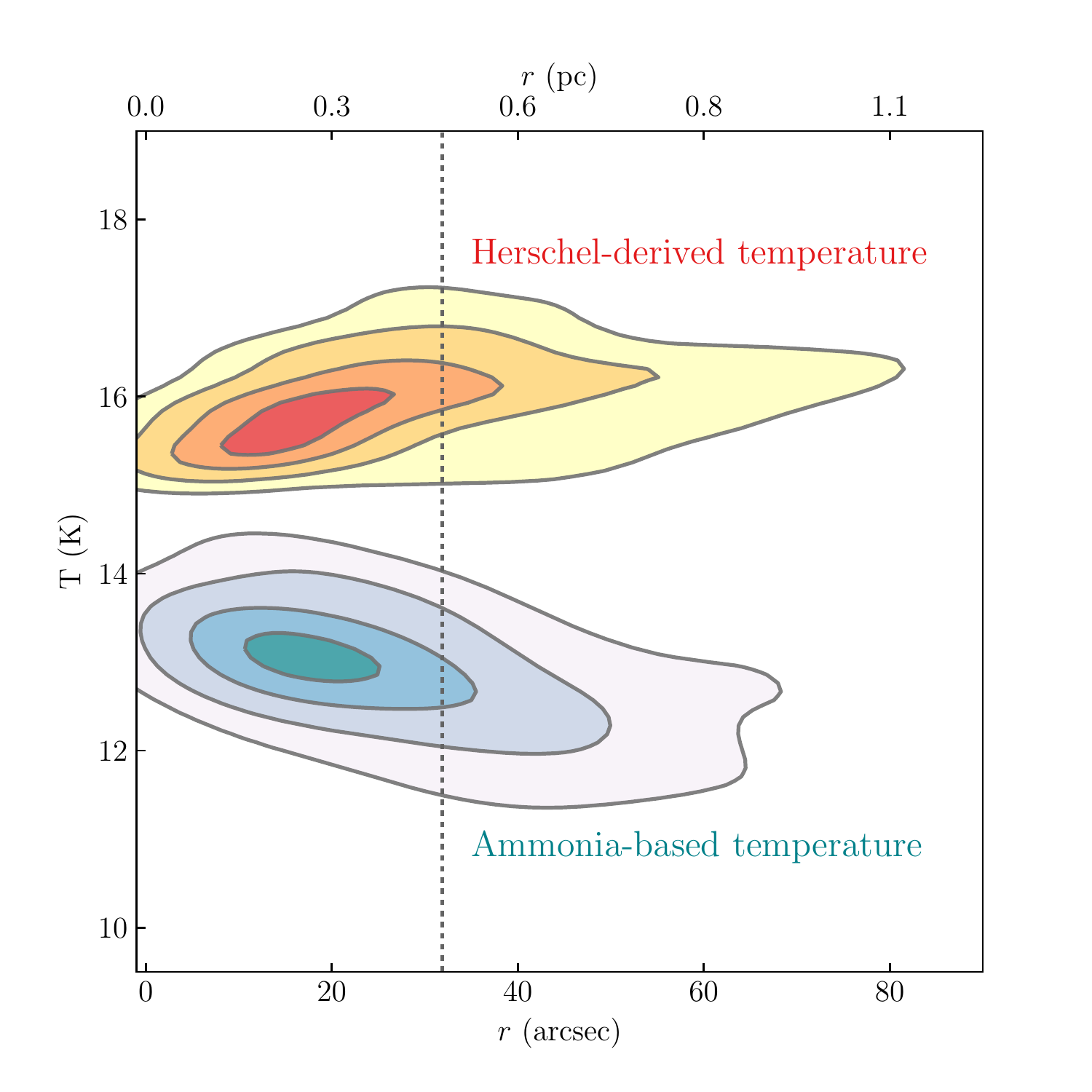}{(a)}{black}
    \subfigcolorrightlabelnew{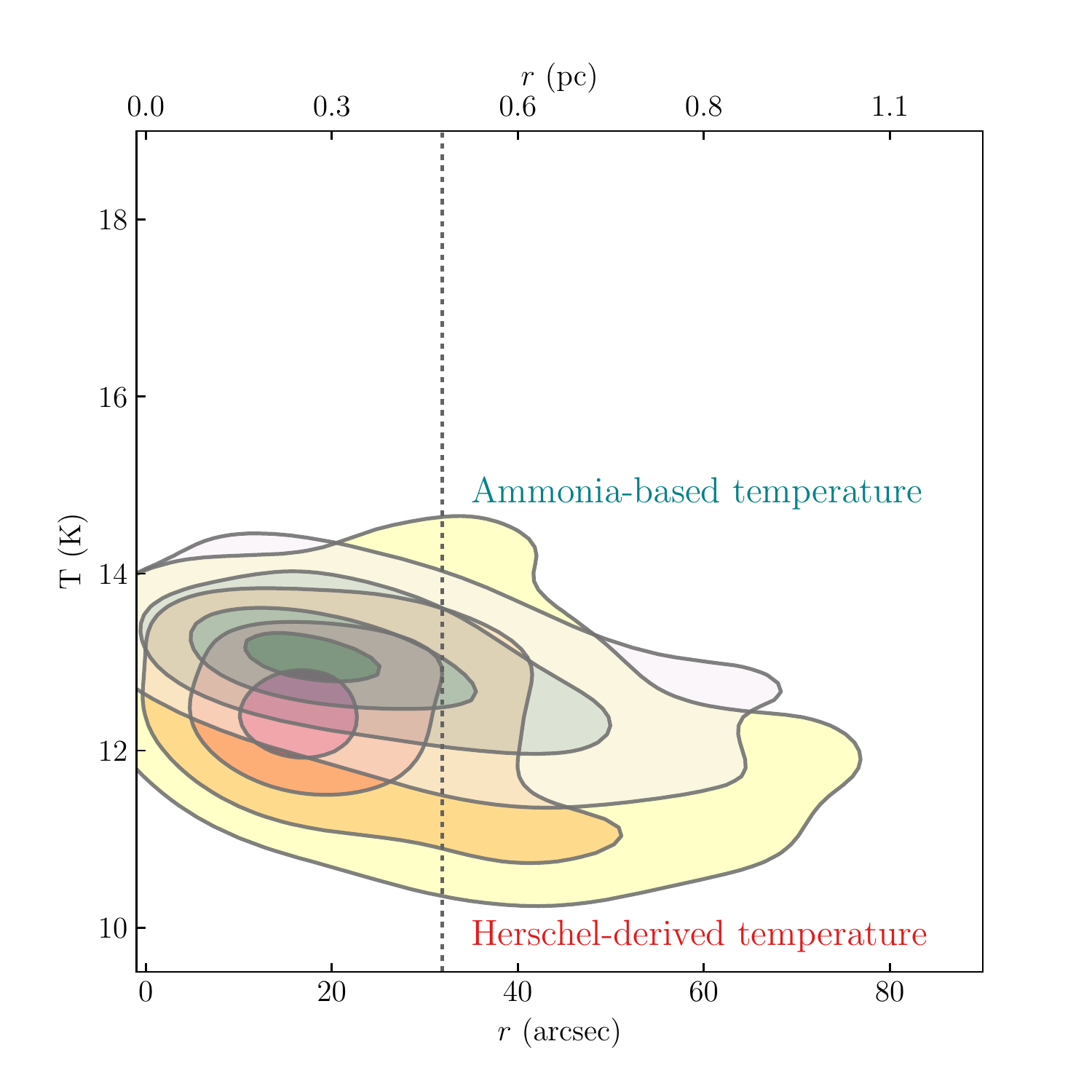}{(b)}{black}
    \caption{(a): Fitted temperatures as a function of distance from the closest massive dense core. The contour overlays show varying levels of kernel density estimation of dust (in red) and gas (in blue) temperatures. The contour $\sigma$-levels are equivalent to that of a bivariate normal distribution, starting at $0.5\sigma$ and progressing outwards in steps of $0.5\sigma$. The vertical dashed line indicates the beam size of the GBT. (b): same as panel (a), but for the dust temperatures derived via the Galactic Gaussian (GG) method.}
    \label{fig:dist_vs_T}
\end{figure}

While the \herschel dust temperature map does not show any signs of local enhancements in the vicinity of dense cores, our \nhhh results point to local temperature increases of about 1 K toward the sites of active star formation. The magnitude of the effect is comparable to what \cite{foster+2009} find towards Perseus, where protostellar cores have been found to have ammonia-derived kinetic temperatures 1.3 K larger than the starless cores.
To further investigate the possibility of dense gas heated by embedded protostars, we split the pixels in two groups, those within one GBT beam from the massive dense cores in \cite{nguyen_luong+2011} and those outside of it. A number of cores with mass lower than $20~\mathrm{M_{\odot}}$ were identified by \cite{nguyen_luong+2011} in addition to MDCs. As the total luminosity output of those cores is, on average, half of that of the MDCs, we restrict the spatial temperature distribution analysis to distance from the MDCs only. We find the median values of the gas kinetic temperatures to be 13.1 K inside the one beam radii, and 12.5 K outside of them. On the contrary, the median dust temperature increases from 15.9 to more than 18 K away from the massive dense cores (Fig. \ref{fig:higal_TN}).
In order to visualize the opposite trends in gas and dust temperatures, we plot their kernel density estimations as a function of distance to the closest massive \herschel core (Fig. \ref{fig:dist_vs_T}a).
The contours on Figure \ref{fig:dist_vs_T} represent the levels of the density function based on the scatter of the temperature points only, as the fitting uncertainties in both ammonia- and dust-based temperatures are much smaller than the dynamic range of the trends in the figure. The mean temperature errors and their standard deviations are $\sigma_{\mathrm{T_{kin}}} = 0.38 \pm 0.15$ K and $\sigma_{\mathrm{T_{dust}}} = 0.36 \pm 0.09$ K for gas and \herschel dust temperatures, respectively.

Systematically different values of dust and gas temperatures stress the importance of considering the line of sight mixing of the IRDC component with the warm dust emission.
A number of methods for estimating the background component in \herschel photometric data is available \citep[e.g.,][]{wang+2015}\footnote{\url{https://github.com/esoPanda/FTbg}}.
To address the line of sight contamination effects on the dust temperatures, we model the line of sight contamination of dust emission via two approaches:
\begin{itemize}
    \item A Galactic Gaussian method \citep[GG,][]{battersby+2011}, assuming that the Galactic contribution follows a Gaussian profile along the latitude direction within $-1\deg \leq b \leq +1\deg$ latitude interval;
    \item A Small Median Filter method (SMF), which interpolates the background contribution from the region outside a predefined IRDC boundary \citep{simon+2006, butler+tan2009} and estimates the 160 to 500\um foreground by utilizing SED model of diffuse ISM \citep{draine+li2007} normalized to the available 24\um extinction data \citep{lim+2014, lim+2015}.
\end{itemize}
A detailed description of each method is presented in \cite{lim+2016}.
The GG- and SMF-processed \herschel maps were used to derive the corrected column density and dust temperature maps following \S\ref{sec:dust}.

We find that both subtraction methods result in lower dust temperatures in G035.39, with the average values for the GG and SMF dust temperatures being $12.4 \pm 1.0$ and $13.8 \pm 0.9$ K, respectively. These values, compared to the mean gas kinetic temperature $T_{kin} = 12.9 \pm 0.8$ K, suggest that the corrections applied to the \herschel maps shifted the peak of the dust SEDs into the temperature regime that much better reflects the actual gas temperature.
Figure \ref{fig:dist_vs_T}b shows the same distance-temperature relation as Fig. \ref{fig:dist_vs_T}a, but for the GG method, a method that matches the ammonia temperature trend the closest. Despite the two trends showing some degree of agreement, no significant correlation is found between the two temperatures (Pearson's $r=0.26$).

\subsection{Stability of the filament}
\label{sec:stability}
The support of a filament against the gravitational collapse is often discussed in terms of its mass to length ratio, or a line mass $(M/L)$. We estimate the line mass from the Hi-GAL $\mathrm{H_2}$ column density map on the pixels that have ammonia detection. The total mass can be estimated as a sum over the column density pixels as follows:
$$ M = \mu_{\mathrm{H_2}} m_{\mathrm{H}} D^2 \int{N_{\mathrm{H_2}} d{\Omega}}, $$
where $D = 2.9$ kpc is a kinematic distance to G035.39.

The mass estimate above is representing a sum over all optically thin dust emission, including both the physical region traced by the GBT ammonia observations and a contribution along the line of sight.
To subtract line of sight contamination, we
we make a simplistic assumption that the total gas column of this extra LoS material is equal to a mean $N{_{\mathrm{H_2}}}$ value along the bounding contour around the \nhhh (2, 2) detection of the main velocity component.
We apply a correction offset of $2.64 \times 10^{22}$ cm$^{-2}$ to the G035.39 gas column density map.
By summing over the column densities within the bounding contour, we find the mass of G035.39 to be $1218$ \Ms, corresponding to $(M/L) \approx 223$ \Mspc.
We argue that this value, limited both by the sensitivity-limited (2, 2) detection and the column density correction above, should resemble a lower limit to the line mass of the filament. Similarly derived (M/L) value for the (1, 1) detection yields the line mass of $319$ \Mspc, while the mass to length ratio of the filament without the envelope subtraction is $634$ \Mspc.
Similarly, if the GG and SMF-derived densities are used to calculate the line mass as above, we obtain (M/L) values of $635$ and $494$ \Mspc, respectively, for slightly different values of $\mathrm{T_{dust}}$.
The representative range of line mass values, $223-635$ \Mspc, is an order of magnitude higher than the critical line mass of thermally supported filament at 15 K, highlighting the importance of other means of support, such as non-thermal motions or magnetic fields.

Given our ammonia fitting results, we can calculate what critical line mass would be needed for the filament to be in equilibrium.
By adapting the conventional filamentary virial analysis \citep{chandrasekhar+fermi1953, ostriker1964} to include both thermal and non-thermal support \citep{fiege+pudritz2000} one can estimate the critical line mass to be
$$ \left(\frac{M}{L}\right)_{crit} = \frac{2 \sigma_{tot}^2}{G}, $$
where $\sigma_{tot}^2 = \sigma_{nt}^2 + c_s^2$ is a quadrature sum of the non-thermal velocity dispersion $\sigma_{nt}$ and the isothermal sound speed $c_s$.
The range of non-thermal motions representative of the filament points
to the critical line masses from 50 to 200 \Mspc, consistently lower than the \herschel-derived (M/L) range. Similarly, this critical line mass value regime is lower than the $(M/L)_{crit}$ values derived from the CO emission \citep[470-1070 \Mspc,][]{hernandez+tan2011, hernandez+2012}. This apparent disparity may result from the bulk of ammonia emission being more sensitive to the inner filament material of G035.39 than the CO data, as the former is known to be relatively enhanced in the denser regions of starless cores \citep{tafalla+2002}, where the CO is frozen-out onto dust grains. This proposition is supported by PdBI observations of continuum structures with much smaller spatial scales (0.03-0.07 pc), where \cite{henshaw+2016_pdbi} suggest that magnetic fields are playing an important role against gravitational collapse of compact continuum cores. We suggest that their conclusions on the dynamical state of compact dense cores may be extended to the larger scale ammonia filament in this study.

\section{Conclusions}
\label{sec:conclusions}
IRDC G035.39--00.33 was studied in its entirety with the Green Bank Telescope. The morphology of the cloud at this resolution resembles that of a smooth filament, more than six parsec in projected length. Multiple distinct components are present along the line of sight throughout the length of the cloud.
\begin{enumerate}
    \item We derive a reliable map of the gas temperature for the entire G035.39 cloud. The extended gas reservoir in the IRDC is consistent with being of starless nature (\Tkin$\sim 11-15$ K). We find evidence of gas heating from the embedded protostars, manifested as slight temperature increase around the positions of 70\micron sources.
    \item Despite having a similar angular resolution, the \herschel observations point to temperatures 2-3 K higher than that of the GBT observations, which can be attributed to the effects of line of sight contamination. We show that this offset can largely be mitigated by accounting for the background and foreground emission components. Reliance on the dust temperature maps derived from conventional FIR SED fitting with no accounting for the line of sight contributions may consistently overestimate the temperature of the dense gas.
    \item We find the brightest velocity component of G035.39 to exhibit a smooth, consistent velocity gradient of ${\sim}0.2$ \kmspc magnitude. This velocity gradient is much lower than that found on smaller scales towards G035.39-N, but is consistent with the global gas motions on parsec and GMC scales.
    \item Additionally to the large-scale gradient across the filament, local velocity field irregularities point to a presence of substructure and smaller-scale fragmentation at scales not traced by the GBT. We suggest that an intertwined network of compact filaments might exist in the southern part of the IRDC, possibly continuing from the one found in G035.39-N. Higher angular resolution observations are needed to fully resolve the sub-structured kinematics of the cloud.
\end{enumerate}

\begin{acknowledgements}
We would like to thank the anonymous referee for the comments which helped to improve this manuscript.
The Green Bank Observatory is a facility of the National Science Foundation operated under cooperative agreement with Associated Universities, Inc.
VS, JEP, and PC acknowledge the support from the European Research Council (ERC; project PALs 320620).
KW is supported by grant WA3628-1/1 of the German Research Foundation (DFG) through the priority program 1573 ("Physics of the Interstellar Medium").
IJ-S acknowledges the financial support received from the STFC through an Ernest Rutherford Fellowship (proposal number ST/L004801/2).
JCT acknowledges NASA grant 14-ADAP14-0135.
This research made use of Astropy, a community-developed core Python package for Astronomy \citep{astropy}, and of APLpy, an open-source plotting package for Python \citep{aplpy}. In addition, this work made use of the \texttt{dust\_emissivity} package\footnote{\url{https://github.com/keflavich/dust\_emissivity}} in the derivation of \herschel dust properties.
\end{acknowledgements}

\bibliography{g35-nh3-arxiv}

\end{document}